%% file: main.tex
\documentclass[a4paper,11pt]{article}
\usepackage{tabularx} 
\usepackage{amsmath}  
\usepackage{amssymb}  
\usepackage{graphicx} 
\usepackage[margin=1in,letterpaper]{geometry} 
\usepackage[final]{hyperref} 

\hypersetup{
	colorlinks=false,       
	linkcolor=blue,        
	citecolor=blue,        
	filecolor=magenta,     
	urlcolor=blue         
}

\usepackage[affil-it]{authblk}  

\usepackage[font=footnotesize,labelfont=bf]{caption} 

\usepackage{physics} 
\usepackage{slashed} 

\usepackage{xcolor} 
\usepackage{xfrac}

\usepackage{float} 
\usepackage{multirow}
\usepackage{makecell}
\usepackage{subcaption}
\usepackage{wrapfig}
\usepackage{enumitem}

\usepackage{cellspace}
\usepackage{bbold}
\usepackage{bm}
\usepackage{dsfont}
\usepackage[normalem]{ulem}
\usepackage{mathrsfs}
\DeclareMathAlphabet{\bscr}{U}{BOONDOX-calo}{m}{n}

\usepackage{titlesec}
\titleformat*{\subsection}{\bfseries\boldmath}
\titleformat*{\subsubsection}{\bfseries\boldmath}

\newcommand{\M}{\mathcal{M}}
\renewcommand{\O}{\mathcal{O}}

\newcommand{\PL}{\text{P}_\text{L}}
\newcommand{\PR}{\text{P}_\text{R}}

\newcommand{\Retilde}{\widetilde{\Re}}

\newcommand{\MSbar}{\ifmmode
    \overline{\text{MS}}
  \else
    $\overline{\text{MS}}$
  \fi}
\newcommand{\lambar}{\bar\lambda}
\newcommand{\lambdamax}{\ifmmode
    \lambda_\text{max}
  \else
    $\lambda_\text{max}$
  \fi}

\newcommand{\amu}{a_\mu}

\newcommand{\SI}[2]{\ifmmode
    #1 \, \text{#2}
  \else
    $#1 \, \text{#2}$
  \fi}

\newsavebox{\figbox}

\usepackage[style=numeric-comp, backend=bibtex, sorting=none]{biblatex}
\title{\bfseries\boldmath On-shell Mass Renormalisation and RG-improved Effective Potential in Vector-like Lepton Models}

\author[1]{Gregor Daberstiel \thanks{gregor.daberstiel@mailbox.tu-dresden.de}}
\author[1]{Kilian Möhling \thanks{kilian.moehling@tu-dresden.de}}
\author[1]{Dominik Stöckinger \thanks{dominik.stoeckinger@tu-dresden.de}} 
\author[1]{Hyejung Stöckinger-Kim \thanks{hyejung.stoeckinger-kim@tu-dresden.de}} 
\affil[1]{Institut für Kern- und Teilchenphysik, TU Dresden, Zellescher Weg 19, 01069 Dresden, Germany}
\date{}

\begin{document}
\maketitle


\begin{abstract}
  \noindent
	We discuss the on-shell mass renormalisation in a class of six vector-like lepton (VLL) extensions of the Standard Model
	with seesaw-like contributions to the SM lepton masses.
	The schemes absorb large radiative corrections to the SM lepton masses arising from mixing between
	the SM and VLLs with up to $5\times5$ mixing matrices and allows for efficient computation of
        observables and the low-scale \MSbar values
	of model parameters. We determine and solve the redundancy-free
        2-loop renormalisation-group (RG) equations, compute the RG-improved
        effective potential 
	and study the vacuum instability scales of each of the VLL models. 
	We find rich phase behaviour and parameter regions that  show instability scales as low as \SI{10^5}{GeV}as well as regions where the one-loop corrections to the input parameters can lead to a stable vacuum.
\end{abstract}

\setcounter{tocdepth}{2}
\tableofcontents

\input{Introduction}
\input{Renormalization}
\input{Potential}
\input{Conclusions}

\section*{Acknowledgments}
We acknowledge financial support by the German Research Foundation
(DFG) under grant number STO 876/7-2 and STO 876/10-1.
\input{Appendix}

\printbibliography
	
\end{document}

%% file: Introduction.tex
\section{Introduction}

Extensions of the Standard Model (SM) by vector-like fermions provide
attractive alternatives to extensions of the scalar sector such as the
Two-Higgs doublet model (2HDM). Though vector-like fermions do not
change the tree-level Higgs potential, they can strongly alter the
fermion mass generation mechanism, and via loops also the effective Higgs potential. Their study can shed light on
the origin of mass and electroweak symmetry breaking.
A special class are vector-like lepton (VLL)
models with seesaw-like mass contributions \cite{delAguila:2008pw,Kannike:2011ng}. Here, the observed lepton
masses arise via mixing with heavy states, and large chirally enhanced
effects on low-energy lepton observables, electroweak and Higgs
observables are possible
\cite{delAguila:2008pw,Kannike:2011ng,Dermisek:2013gta,Falkowski:2013jya,Guedes:2022cfy,Hamaguchi:2022byw,Dermisek:2022aec,Dermisek:2023nhe,Mohling:2024qvk,Daberstiel:2026wxe}.\footnote{For
more general investigations of vector-like fermions we refer to Refs.~\cite{Langacker:1988ur,delAguila:1989rq,Dimopoulos:1990kc,Fujikawa:1994we,Sher:1995tc,Thomas:1998wy,Frampton:1999xi,delAguila:2008pw,
	Ishiwata:2011hr,Kannike:2011ng,Buras:2011ph,Joglekar:2012vc,Kearney:2012zi,Dermisek:2013gta,Ishiwata:2013gma,Falkowski:2013jya,Freitas:2014pua,
	Ellis:2014dza,Ishiwata:2015cga,Poh:2017tfo,Patel:2017pct,Alonso:2018bcg,Endo:2020tkb,Crivellin:2020ebi,Bhattacharya:2021ltd,Cherchiglia:2021syq,Jana:2021tlx,Baspehlivan:2022qet,Dermisek:2022aec,Dermisek:2023nhe,Arkani-Hamed:2021xlp,Hamaguchi:2022byw,deBlas:2025pco,Erdelyi:2025axy,Arkani-Hamed:2026wwy}.}

Higher-order corrections and renormalisation are often required in
beyond-the-SM (BSM) studies, especially in the current era of
high-precision measurements and overall excellent agreement of
experiments with SM predictions. An example in the context of VLL
models is provided by Ref.\ \cite{Mohling:2024qvk}, where a specific
VLL scenario with ``opposite sign'' Yukawa coupling was considered. In
this scenario, the measured Higgs--muon coupling implies a {\em lower}
limit on the VLL contributions to the anomalous magnetic moment of the
muon $\amu$ in conflict with current $\amu$-constraints. However, loop
corrections can significantly lower  this limit \cite{Mohling:2024qvk,Mohling:2026fqu}. 
Similarly, the loop corrections can also lead to strong constraints on the stability of the vacuum
 \cite{Frampton:1999xi,Joglekar:2012vc,Gopalakrishna:2018uxn,Arsenault:2022xty,Cingiloglu:2023ylm,Adhikary:2024esf,Cingiloglu:2024vdh,Bednyakov:2025uur}.

The construction of renormalisation schemes in many extensions of the
SM is significantly more complicated than in the SM itself. The SM
(without generation mixing) has the special property that all input
parameters except the electromagnetic and strong gauge couplings
are masses of physical particles that can be on-shell
renormalised; conversely, all physical masses are free parameters. In
many BSM models, one or both statements become invalid.

For example, in the Minimal Supersymmetric Standard Model (MSSM) or the 2HDM, there appear
important parameters such as $\tan\beta$ or other mixing angles for
which neither an on-shell condition nor a direct relationship to an
observable is straightforward. Criteria such as numerical stability,
gauge independence and process independence have been put forward to
analyse such schemes \cite{Freitas:2002um,Denner:2018opp}, but it is usually not possible to define
and renormalise such parameters in a problem-free way
\cite{Santos:1996vt,Freitas:2002um,Heinemeyer:2010mm,Krause:2016oke,Altenkamp:2017ldc,Denner:2018opp,Grimus:2018rte,Dudenas:2020ggt,Dittmaier:2022maf,Dittmaier:2022ivi}.
In addition, in supersymmetric models, several masses such as Higgs, neutralino or
squark masses do not correspond to free parameters and thus cannot
be on-shell renormalised \cite{Hollik:2002mv,Heinemeyer:2004gx,Slavich:2020zjv}.
Here we construct full renormalisation schemes for the six simplest VLL models
with seesaw-like mass contributions, allowing for CP violation, and provide the on-shell
conditions and respective solutions to the renormalisation constants (RCs). 
In particular, the SM lepton masses are renormalised on-shell in all schemes. This avoids
the otherwise computationally expensive numerical iteration to determine the tree-level
value of the Yukawa couplings and instead allows us to obtain the \MSbar couplings in a more straightforward way.
We apply this procedure to compute the renormalisation-group (RG) improved effective potential
at next-to-leading log (NLL) accuracy and evaluate the constraints from vacuum stability on the input parameters in the
chosen schemes.
Among the goals and issues we highlight the following: 
\begin{itemize}
\item
  The schemes employ on-shell conditions for physical masses wherever
  possible. However, like e.g.\ in the MSSM neutralino sector, not all
  masses are independent, preventing a complete on-shell renormalisation of all heavy lepton masses.
\item
  Our treatment is similar to the one of the neutralino sector in
  Refs.~\cite{Hollik:2002mv,Fritzsche:2002bi}, but the mixing involves
  up to $5\times5$ matrices. For one of the six VLL models,
  Ref.~\cite{Mohling:2024qvk} has already presented a possible scheme;
  here we present additional schemes with improved numerical stability
  over a wider range of parameters and extend the construction to all six models.
\item
  For the remaining non-mass parameters we choose a suitable set of
  fundamental parameters of the Lagrangian before electroweak symmetry
  breaking and renormalise them in the \MSbar scheme. The following two items are important in this context.
\item
  In order to maximise the numerical stability of perturbation theory
  it is advantageous to make sure that the finite parts of
  renormalisation constants are not too large, especially not enhanced
  by factors such as $m_L/v$, where $m_L$ is a heavy VLL mass and $v$
  the electroweak scale.
\item
  There is a redundancy between fundamental Lagrangian parameters and
  field redefinitions that can be used to eliminate
  certain parameters and optionally also their renormalisation
  constants. Depending on how the redundancy is dealt with, the form of the renormalisation group equations (RGEs) differ. 
\item 
  For the analysis of vacuum stability at NLL accuracy, using the appropriate
  $\beta$-functions at the two-loop level together with the NLO initial values of the \MSbar couplings is crucial.
\end{itemize}

The paper is structured as follows. In Sec.~\ref{sec:Renormalisation} we first briefly summarise the model definitions, distinctive properties
and renormalisation procedure of the gauge and Higgs sectors, which remains the same as in the SM. We then discuss in detail how the renormalisation of
the lepton sector is set up, which on-shell conditions are imposed and how the fundamental RCs are obtained from these conditions. At the end of this 
section we also discuss how to address the parameter redundancy and compute the RG $\beta$-functions that preserve the redundancy-free structure
of the coupling matrices at two-loop order.
In Sec.~\ref{sec:Potential} we discuss the calculation of the effective potential at NLL, in particular including the one-loop corrections from
the vector-like and SM leptons. We explain how the \MSbar Higgs mass parameter is obtained consistently in light of the fine-tuning problem
and compute the scale at which the EW vacuum becomes unstable. We also provide estimates for the uncertainty from missing higher-order corrections
and input values. At the end we provide numerical results as well as approximate formulas that allow us to determine the instability scale for all 6 models.
In Sec.~\ref{sec:Conclusion} we summarise the main results and provide some concluding remarks.

%% file: Renormalization.tex
\section{Renormalisation scheme and on-shell conditions}\label{sec:Renormalisation}

\subsection{Model definitions}
\label{sec:ModelDefinitions}

The six models of interest for the present paper have been
systematically introduced in Ref.~\cite{Kannike:2011ng} and have also been
discussed in Ref.~\cite{Daberstiel:2026wxe}, where
definitions and tree-level formulas for Lagrangians, mass matrices and
Feynman rules were given. We follow the conventions of
Ref.~\cite{Daberstiel:2026wxe} and refer to it for details. Here we
only summarise the definitions 
 briefly.

Each of the six models adds two VLL
fields to the SM. Schematically, the six possibilities are
	\begin{align}\label{chargedseesawpossibilities}
		L\oplus E, \qquad L\oplus N, \qquad L_{\frac{3}{2}}\oplus E, \qquad L\oplus E^a, \qquad L\oplus N^a, \quad \text{and} \quad L_{\frac{3}{2}}\oplus E^a.
	\end{align}
Here each field is a VLL, i.e.\ a Dirac fermion with one unit of
lepton number. $L$, $L_{\frac{3}{2}}$ denote doublets with hypercharge $Y=-1/2$,
$-3/2$; $E$ and $E^a$ are a singlet/triplet with hypercharge $Y=-1$; and
$N$, $N^a$ are a singlet/triplet with $Y=0$. The first three models
contain VLL singlets, and we will refer to them collectively as
singlet models; the last three models will be referred to as triplet
models.

All six models
lead to an analogous structure of BSM terms in the Lagrangian. To
simplify the notation, the parameter names are chosen to be equal in
all models, choosing the letter $L$ for all parameters associated
with VLL doublets and $E$ for all parameters associated
with VLL singlets or triplets. Then, in each of the models, there is
the following set of 13 lepton sector parameters,
\begin{align}\label{leptonsectorparameters}
  M_L, M_E; \lambda^L_i, \lambda^E_i; \lambda, \bar{\lambda}; y_i.
\end{align}
Here, $M_{L,E}$ are the corresponding Dirac mass parameters,
$\lambda^L_i$ are Yukawa couplings of the VLL doublet to the SM
singlets $e_{Ri}$ and the SM Higgs doublet, where $i\in\{1,2,3\}$ is a generation index, $\lambda^E_i$ are
Yukawa couplings involving the VLL singlet/triplet and the SM doublets
$l_{Li}$. The parameters $\lambda$, $\bar{\lambda}$ are Yukawa couplings
involving both BSM fields, where $\lambda$ involves the left-handed
part of the VLL doublet and $\bar{\lambda}$ the right-handed part of
the VLL doublet. Finally, $y_i$ are the three generation-diagonal SM
lepton Yukawa couplings.

The SM gauge and Higgs sector are unchanged, and the SM quark and
lepton fields have the same interactions as in the SM, but the VLL
interactions mentioned above strongly modify the lepton sector. The
major common feature of all six models is the existence of 
chirally enhanced contributions to the lepton masses, lepton--Higgs
couplings and the lepton dipole moments. The leading terms of these
contributions are governed by combinations of the form
\begin{align}
  \frac{\bar{\lambda}\lambda^E\lambda^Lv^2}{M_E M_L},
\end{align}
resulting in strong and model-specific correlations that were studied in
the literature at lowest order 
\cite{Kannike:2011ng,Dermisek:2013gta,Falkowski:2013jya,Hamaguchi:2022byw,Dermisek:2022aec,Dermisek:2023nhe,Daberstiel:2026wxe}
and in special cases taking into account higher orders \cite{Mohling:2024qvk,Athron:2025ets,Mohling:2026fqu}.

In more detail, each model contains 4 or 5 charged ($Q=-1$) leptons, and each model contains 3, 4 or 5 neutral
leptons/neutrinos. In addition, some models contain doubly
charged or positively charged leptons. Table \ref{tab:field-content}
specifies the appearing particles exactly. After electroweak symmetry breaking, off-diagonal mass matrices   $\mathscr{M}_f$ arise for each of the multiplets ($f=e,\nu,\rho,\chi$), requiring transformations to the mass basis in order to obtain diagonal lepton mass terms and physical lepton fields. The explicit form of all the
relevant mixing matrices can be found in the mentioned literature.

\begin{table}[t]
	\centering
	\def\arraystretch{1.2}
	\begin{tabular}{c||c|c|c|c|c|c}
		\textbf{Model} & $L\oplus E$ & $L\oplus N$ & $L_{\frac{3}{2}}\oplus E$ & 
		$L\oplus E^a$ & $L\oplus N^a$ & $L_{\frac{3}{2}}\oplus E^a$ \vspace{.15em} \\ \hline\hline
		$n_\nu$ ($Q=0$)\ & 4 & 5 & 3 & 5 & 5 & 4 \\
		$n_e$  ($Q=-1$)  & 5 & 4 & 5 & 5 & 5 & 5 \\
		$n_\rho$ ($Q=-2$)  & 0 & 0 & 1 & 1 & 0 & 2 \\
		$n_\chi$ ($Q=+1$) & 0 & 0 & 0 & 0 & 1 & 0 \\
	\end{tabular}
	\caption{Number of mass eigenstates for the different electric
          charges $Q$ in the six VLL models.}
	\label{tab:field-content}
\end{table}

\subsection{Renormalisation of the gauge and Higgs sector}

Before discussing the renormalisation of the lepton sector, we briefly collect the relevant relations for the renormalisation constants (RCs)
of the gauge and Higgs sector which are identical to the SM.
In these sectors we use the full on-shell renormalisation scheme for the SM as
defined in Refs.~\cite{Bohm:2001yx,Denner:2019vbn}; this scheme was
applied to VLL models already in Ref.~\cite{Mohling:2024qvk}, and we
follow the notation of this paper. The Higgs Lagrangian is written as
\begin{align}
	\mathcal{L}\supset (D_\mu\Phi)^\dagger(D^\mu\Phi) - V(\Phi) \qquad \text{with} \qquad V(\Phi) = -\mu^2|\Phi|^2 + \lambda_\Phi |\Phi|^4,
\end{align}
and the Higgs vacuum expectation value (\emph{vev}) at the electroweak
minimum is denoted by $\ev{\Phi} = (0,v)^T$ (with $v\approx 174$ GeV).
The Lagrangian is rewritten in terms of mass-eigenstate fields and
appropriate parameters after electroweak symmetry breaking (EWSB);
then a renormalisation transformation is applied and parameters and fields are replaced by bare quantities.
The bare tadpole parameter, fundamental charge and the boson masses are given in terms of the bare potential parameters and gauge couplings by
\begin{subequations}
	\begin{alignat}{2}
		&t_0 = \sqrt{2} v_0 \Big(\mu_0^2 - 2\lambda_{\Phi,0} v_0^2\Big), \qquad && M_{h,0}^2 = 6\lambda_{\Phi,0} v_0^2 - \mu_0^2, \\
		&e_0 = g_{Y,0} \, c_{W,0}, \qquad && M_{W,0}^2 = c_{W,0}^2 M_{Z,0}^2 = \frac{1}{2} g_{W,0}^2 v_0^2,
	\end{alignat}
\end{subequations}
where $c_{W,0} = g_{W,0}/\sqrt{g_{W,0}^2 + g_{Y,0}^2}$. The fundamental charge and boson masses are renormalised by imposing the usual on-shell relations on the
boson self-energies, while the tadpole parameter is set to
$t=0$ at tree-level such that $t_0 = \delta t = - T_h$ subtracts the radiative
Higgs tadpole diagrams $T_h$. In general, all bare quantities are split as
$\Theta_0=\Theta+\delta\Theta$ into a tree-level, or renormalised,
part $\Theta$ plus a RC $\delta\Theta$. In this
way the correct form of the counterterm Lagrangian is generated.
At one-loop order, the RC for the Higgs \emph{vev} can be written as
\begin{align}\label{eq:dv}
	\frac{\delta v}{v} = \frac{\delta M_W^2}{2M_W^2} + \frac{c_W^2}{s_W^2}\bigg(\frac{\delta M_Z^2}{2M_Z^2}-\frac{\delta M_W^2}{2M_W^2}\bigg) - \frac{\delta e}{e},
\end{align}
and the RCs for the potential parameters as
\begin{align}\label{eq:potential-parameter-RC}
	\delta \mu^2 = \frac{1}{2}\bigg(\delta M_h^2 + 3 \frac{\delta t}{\sqrt{2}v}\bigg), \qquad
	\delta \lambda_\Phi = \frac{1}{4v^2} \bigg(\delta M_h^2 + \frac{\delta t}{\sqrt2 v} - 2 M_h^2 \frac{\delta v}{v}\bigg).
\end{align}

\subsection{Lepton renormalisation transformation: symmetric
  parameters and
  redundancies}

Here we begin the construction of the renormalisation schemes for the
lepton sectors of all six VLL models. Our general strategy is the same
as the one of Ref.~\cite{Mohling:2024qvk} which in turn follows the
procedure of Refs.~\cite{Hollik:2002mv,Fritzsche:2002bi} for the
MSSM chargino/neutralino sector. We first set up the form of the
counterterm Lagrangian, then specify the on-shell mass and field
renormalisation conditions, and finally we discuss the renormalisation
of other parameters and the practical determination of all
renormalisation constants.

The counterterm Lagrangian itself is generated in two steps. Here we
discuss the first step that starts from the Lagrangian
written in terms of gauge-eigenstate fields and generates RCs for the
corresponding symmetric parameters.

An important complication is the existence of redundancies between
fields and parameters, already discussed in
Ref.~\cite{Mohling:2024qvk} for the $L \oplus E$ model. Such
redundancies exist in all six VLL models, and they imply that certain
Lagrangian parameters are unphysical and can be eliminated via field
redefinitions. Similar issues also
arise e.g.\ in the 2HDM Higgs sector (here, Ref.~\cite{Herren:2017uxn} has pointed out
  that $\beta$ functions of redundant/unphysical parameter
  combinations can be divergent) and the SM CKM sector.

In our case, field redefinitions allow  transforming each left-handed
and right-handed part of each VLL multiplet
in Eq.~(\ref{chargedseesawpossibilities}) by a complex phase, and the
three generations of SM lepton doublets and singlets $l_{Li}$ and
$e_{Ri}$ can be transformed with unitary $3\times3$ matrices
$V_L\in \text{U}(3)$ and $V_R\in \text{U}(3)$. In addition, in the VLL models
containing the doublet $L$ and/or the singlet $E$, one can form the
4-tuples $\bscr{l}_{La}=(l_{Li},L_L)$ and/or $\bscr{e}_{Ra}=(e_{Ri},E_R)$ with
$a=1\ldots4$ and $i=1\ldots3$. These 4-tuples involve only fields with
identical gauge quantum numbers, and the $4\times4$ field
redefinitions
\begin{align}\label{eq:redundancy-transformation}
	\bscr{l}_{La} \to (V_L^\dagger)_{ab} \, \bscr{l}_{Lb} \qquad \text{and/or} \qquad \bscr{e}_{Ra} \to (V_R^\dagger)_{ab} \, \bscr{e}_{Rb},
\end{align}
with $V_L\in \text{U}(4)$ and/or $V_R\in \text{U}(4)$ are possible.

The field redefinitions can be used to eliminate unphysical
off-diagonal Yukawa couplings and mass terms. Ultimately, in all six
VLL models, the set of 13 physical lepton sector parameters already listed in
Sec.~\ref{sec:ModelDefinitions}, Eq.~(\ref{leptonsectorparameters}) emerges.
The freedom of phase transformations could be used to absorb unphysical
complex phases, such that w.l.o.g.\ only 13 positive real parameters
$|y_i|,|\lambda^{L,E}_i|,|\lambda|,|\lambar|$ and $|M_{E,L}|$ 
and 4 physical CP violating phases (e.g.\  $\arg y_i$ and $\arg
\lambar$) remain. For the purposes of the present paper we find it
convenient to eliminate the ``off-diagonal'' parameters but to keep
the redundant phases.

Having fixed the parameterisation, the renormalisation transformation
is applied onto all parameters, i.e.\ the bare leptonic Lagrangian is
generated from the tree-level one by replacing the 13 parameters by bare
parameters as
\begin{align}\label{eq:fundamental-RC}
	\Theta\to\Theta_0 = \Theta + \delta\Theta \qquad \text{for} \qquad \Theta \in \{y_i, \lambda^{L,E}_{i}, \lambda, \lambar, M_{L,E}\}.
\end{align}
Here the redundancy on the bare level is eliminated in the same way as
on the tree level. The one-loop counterterm Lagrangian is generated by
expanding the bare Lagrangian at first order in the RCs $\delta
\Theta$. For later convenience we also introduce the Laurent coefficients in the $\epsilon$-expansion
as
\begin{align}
	\delta\Theta = \sum_{n=-\infty}^\infty \frac{\delta^{[n]}\Theta}{\epsilon^n},
\end{align}
i.e.\ in particular the finite part of the RCs is denoted by $\delta^{[0]}\Theta$ and \MSbar renormalised
parameters have $\delta^{[n]}\Theta=0$ for $n\leq 0$.

\subsection{Lepton renormalisation transformation: mass-eigenstate fields and counterterms}

The second step in constructing the lepton counterterm Lagrangian is the
introduction and renormalisation of mass-eigenstate fields.

At tree-level, the gauge-eigenstate lepton fields are replaced by
mass-eigenstate  fields defined via unitary matrices $U_{L,R}^f$ that diagonalise
the lepton mass matrices after EWSB,
\begin{align}
	\Big[U_L^{f\dagger} \mathscr{M}_f U_R^f\Big]_{ab} = m_{f_a} \delta_{ab},
\end{align}
where $f=\nu,e,\rho$ and $\chi$. The mass matrices $\mathscr{M}_f$
are explicitly listed in Eq.~(4a-f) of Ref.~\cite{Daberstiel:2026wxe}.
We only emphasise again that they consist of the fundamental parameters listed in
Eq.~\eqref{leptonsectorparameters} and the \emph{vev} $v$, and that they can be split up as
\begin{align}\label{Massmatrixdecomposition}
  \mathscr{M}_f&=
   \mathcal{Y}^h_f \, v +  \mathcal{M}_f^D,
\end{align}
where $\mathcal{Y}^h$ are matrices containing only the dimensionless
Yukawa couplings of Eq.~\eqref{leptonsectorparameters} while
$\mathcal{M}_f^D$ only contain the dimensionful Dirac mass parameters
of Eq.~\eqref{leptonsectorparameters}.

In our approach, the diagonalisation matrices are not renormalised but kept fixed at tree-level, whereas
the fundamental parameters appearing in the mass matrices
$\mathscr{M}_f$ undergo the renormalisation transformations of Eq.~\eqref{eq:fundamental-RC} together with the \emph{vev} renormalisation from Eq.~\eqref{eq:dv}.
After introducing mass-eigenstate fields, the renormalisation
transformations applied to Eq.~(\ref{Massmatrixdecomposition}) lead to
the one-loop mass RC  
\begin{align}\label{eq:mass-basis-mass-RC}
	(\delta m_{f})_{ab} = \Big[U_L^{f\dagger} \delta \mathscr{M}_f
    U_R^f\Big]_{ab} \equiv   \Big[U_L^{f\dagger}
    \Big( \mathcal{Y}^h_f \, \delta v  + v \, \delta \mathcal{Y}^h_f + \delta \mathcal{M}_f^D \Big) U_R^f\Big]_{ab},
\end{align}
that are in general non-diagonal. The gauge-eigenstate
Yukawa and Dirac mass counterterm matrices  $\delta
\mathcal{Y}^h_f$ and $\delta \mathcal{M}_f^D$ are listed for the different
models 
in App.\ref{app:RC-mat}.

Finally, we introduce the arbitrary (complex) field renormalisation
matrices independently for the left- and right-handed mass-eigenstate
lepton fields as
\begin{align}
	f^a_{L/R} \to \sqrt{Z_{L/R}^{f}}_{ab} \, f_{L/R}^b = \Big(\mathds{1} + \tfrac{1}{2}\delta Z^f_{L/R} + \O(\hbar^2) \Big)_{ab} \, f^b_{L/R},
\end{align}
and denote the hermitian/ anti-hermitian parts by $Z_{L/R}^{f,H/A}
\equiv \frac{1}{2}(Z_{L/R}^f\pm Z_{L/R}^{f\dagger})$. This
transformation generates
the final form of the lepton sector counterterm Lagrangian.

\subsection{On-shell conditions on lepton self-energies}
Here we specify the form of the on-shell renormalisation conditions in
the lepton sector. In brief, we require full on-shell conditions for
field renormalisation and on-shell mass conditions for a suitable
subset of (charged or neutral) lepton masses.

In general, we decompose a renormalised fermion self energy into
covariants as
\begin{align}
	\hat{\Sigma}^f_{ab}(p) = \hat{\Sigma}_{ab}^{f,R}(p^2) \slashed{p}\PR + \hat{\Sigma}_{ab}^{f,L}(p^2) \slashed{p}\PL + 
	\hat{\Sigma}_{ab}^{f,SR}(p^2) \PR + \hat{\Sigma}_{ab}^{f,SL}(p^2)\PL,
\end{align}
where $\PL, \PR $ are the left- and right-handed projection
operators. At one-loop order the covariants 
are given in terms of the unrenormalised covariants and
the RCs by
\begin{subequations}\label{eq:SE-renormalised-covariants}
	\begin{align}
		\hat{\Sigma}^{f,L/R}_{ab}(p^2) &= \Sigma^{f,L/R}_{ab}(p^2) + (\delta Z^{f,H}_{L/R})_{ab}, \\
		\hat{\Sigma}^{f,SR}_{ab}(p^2) &= \Sigma^{f,SR}_{ab}(p^2) - \tfrac{1}{2} m_{f_a} (\delta Z_{R}^f)_{ab} 
		- \tfrac{1}{2} (\delta Z_{L}^{f})_{ba}^* m_{f_b} - (\delta m_f)_{ab}, \\
		\hat{\Sigma}^{f,SL}_{ab}(p^2) &= \Sigma^{f,SL}_{ab}(p^2) - \tfrac{1}{2} m_{f_a} (\delta Z^f_{L})_{ab} 
		- \tfrac{1}{2} (\delta Z_{R}^{f})^*_{ba} m_{f_b} -
                (\delta m_f)^*_{ba} .
	\end{align}
\end{subequations}
The physical lepton masses are characterised as poles $p^2=(m^{\text{pole}})^2-im^{\text{pole}}\Gamma$ of the full
renormalised two-point function as a function of the momentum $p^2$
(for unstable leptons the pole has an imaginary part corresponding to
the width). Equivalently, the pole masses satisfy
\begin{align}\label{eq:complex-pole-conditiongeneral}
  \det{(\slashed{p} - m_{f_a}) \delta_{ab}
    + \hat{\Sigma}^f_{ab}(p)}\Big|_{p^2=(m^{\text{pole}})^2-im^{\text{pole}}\Gamma} = 0.
\end{align}
In general, the pole mass of a particle $f_a$ can differ from the
corresponding tree-level mass, $m_{f_a}^{\text{pole}}\neq m_{f_a}$.
An on-shell condition for a tree-level lepton mass $m_{f_a}$ amounts
to the requirement that  the (real part of the) physical 
pole mass is given by the tree-level mass, $m_{f_a}^{\text{pole}} = m_{f_a}$, at all orders. Equivalently,
$p^2=m_{f_a}^2 - im_{f_a}\Gamma_{f_a}$ must solve
Eq.~(\ref{eq:complex-pole-conditiongeneral}) at all orders for some
value of $\Gamma_{f_a}$.\footnote{%
In Ref.~\cite{Mohling:2024qvk} a slightly simpler version of the
on-shell mass condition was used that corresponds to requiring
Eq.~(\ref{eq:no-mixing-condition}) for $a\neq b$ and for $a=b$.
In the review
Ref.~\cite{Denner:2019vbn}, a mass defined as in
Ref.~\cite{Mohling:2024qvk} is called on-shell mass, while a mass
defined via Eq.~(\ref{eq:complex-pole-conditiongeneral}) is called
pole mass. At the
one-loop level considered here, the two conditions are equivalent but
beyond the one-loop level, they differ and the pole mass definition has
conceptual advantages. For our purposes, working
with  Eq.~(\ref{eq:complex-pole-conditiongeneral}) has practical
advantages mainly in the case of CP violation where it more clearly
isolates the relevant RCs for fixing the physical mass.
}
The width  $\Gamma_{f_a}$ is then a
prediction of the theory.

Requiring such an on-shell mass condition at one-loop order is
equivalent to the following expression for the corresponding diagonal mass RC,
\begin{align}\label{eq:mass-RC}
	\Re(\delta m_f)_{aa} &= \frac{1}{2} \Re \Big(\Sigma^{f,SR}_{aa} + \Sigma^{f,SL}_{aa} + m_{f_a} \Sigma^{f,R}_{aa} + m_{f_a} \Sigma^{f,L}_{aa}\Big)_{p^2=m_{f_a}^2}.
\end{align}

Later, it will turn out to be impossible to require such on-shell mass
conditions for all lepton masses in all VLL models. For any
non-on-shell renormalised mass the pole mass $m_{f_a}^{\text{pole}}$
is then a prediction of the theory that may differ from $m_{f_a}$ and
Eq.~(\ref{eq:mass-RC}) is not valid.  

On-shell field renormalisation conditions, however, are required and possible for all
leptons. If $p^2$ is chosen to correspond to the (real part of the) pole mass of lepton
$f_b$, then we first require the no-mixing condition on all
off-diagonal self-energies,
\begin{align}\label{eq:no-mixing-condition}
	\Retilde\ \hat{\Sigma}_{ab}(p) u_b(p) = 0 \qquad \text{for} \qquad a\neq b,
\end{align}
where $\Retilde$ removes the imaginary part of the loop integrals.

Finally, the usual residue condition
\begin{align}\label{eq:residue-condition}
	\lim_{p^2\to m_{f_a}^2} \big(\slashed{p} - m_{f_a}\big)^{-1} \Retilde\ \hat{\Sigma}_{aa}(p) u_a(p) = 0,
\end{align}
is imposed on the diagonal self-energies.

The field renormalisation conditions determine the field RCs as follows,
\begin{subequations}\label{eq:OS-solution}
	\begin{align}
		\delta Z_{L,aa}^{f,H} &= - \Retilde \Big[\Sigma^{L}_f + m_f \Big(\Sigma^{SR\,\prime}_f+\Sigma^{SL\,\prime}_f\Big)
		+ m_f^2
                \Big(\Sigma^{R\,\prime}_f+\Sigma^{L\,\prime}_f\Big)\Big]_{aa}^{p^2=m_a^2}
                , \\
		\begin{split}
			\delta Z_{L,ab}^f &= \tfrac{2}{m_{f_a}^2 - m_{f_b}^2} \Retilde\Big[ \Big(\Sigma^{SR}_f + \Sigma^{L}_f m_f - \delta m_f\Big) m_f
			+ m_f \Big(\Sigma^{SL}_f + \Sigma^{R}_f m_f -
                        \delta m^\dagger_f\Big) \Big]_{ab}^{p^2=m_b^2} ,
		\end{split} \\
		\delta Z_{R,ab}^{f} &= \delta Z_{L,ab}^{f}\{ L\leftrightarrow
                R,~\delta m^\dagger \leftrightarrow \delta m \} .
	\end{align}
\end{subequations}
The equations hold for $a\neq b$ and thus, the diagonal hermitian parts $\delta Z^H_{L/R,aa}$ and the full
off-diagonal RCs $\delta Z_{L/R,ab}$ are determined by the renormalisation conditions on the fermion self-energies. 
Note that the diagonal anti-hermitian (i.e.\ imaginary) part of the field RCs is not determined by the OS conditions
and will instead be fixed below via the redundancy transformation of the parameter RCs.

An exception to the above equations is given by the massless SM-like neutrinos with
$m_{\nu_a} = 0$ for $a=1,2,3$. For these, the off-diagonal field RCs are not fixed by the on-shell conditions and we instead choose them as
\begin{align}
	\delta Z_{L/R, ab}^{\nu,H} = -\Sigma^{\nu,L/R}_{ab}(0),
\end{align}
such that the renormalised covariants of the massless neutrino self-energies vanish at $p^2=0$.

\subsection{Model-specific scheme definitions and solutions for counterterms}

The previous subsections have described how to generate the form of
the counterterm Lagrangians and how on-shell conditions can be solved
for combinations of RCs. We are now in the position to specify
renormalisation schemes for the six VLL models. As a preview, for each
model we will define three schemes that have slightly different
advantages in practical applications. All schemes will impose on-shell
mass conditions on all SM-like lepton masses and a selection of BSM
lepton masses, and $\overline{\text{MS}}$ conditions on a selection of
fundamental parameters.

In the context of dimensional regularisation with $D=4-2\epsilon$, the ultraviolet (UV) divergent parts of the RCs can all be obtained from pure
$\overline{\text{MS}}$ renormalisation and from known
renormalisation-group $\beta$-functions (see the following
Sec.~\ref{sec:RGE}). Furthermore, the field renormalisation constants
(including their finite parts) are fixed by the on-shell field
renormalisation conditions in Eqs.~(\ref{eq:no-mixing-condition},
\ref{eq:residue-condition}). Hence the definition of a scheme only
needs to specify the determination of the finite
parts of the parameter RCs.

Ref.~\cite{Mohling:2024qvk} has shown that the $L\oplus E$ model
admits the following renormalisation scheme: six masses
(all five charged lepton masses and the single heavy neutrino mass)
are on-shell renormalised, the corresponding on-shell conditions directly
fix the corresponding mass counterterms $\bm{\delta}\mathbf{m}
=\{\Re(\delta m_e)_{11}, \ldots, \Re(\delta m_e)_{55},\Re(\delta m_\nu)_{44}\}$ and
can be solved to unambiguously fix the six fundamental RCs
$\bm{\delta}\mathbf{M}=\{v \delta y_i, v\delta\lambda, \delta M_E,
\delta M_L\}$, while the remaining seven fundamental RCs $\delta\lambda^L_i,
\delta\lambda^E_i, \delta\bar{\lambda}$ are
$\overline{\text{MS}}$-renormalised. All subsequent schemes follow the
same idea.

We begin by defining schemes denoted by ``OS1'' for all six
models. These schemes are directly analogous to the scheme of
Ref.~\cite{Mohling:2024qvk} and impose on-shell mass conditions on as
many leptons as possible. In all models involving singlets, there are
three additional heavy lepton states (with charges $0, -1$ or
$-2$), and like in the $L\oplus E$ model we require on-shell
mass renormalisation of all SM-like leptons and all heavy leptons. Hence in each
singlet model a set of six mass counterterms $\bm{\delta}\mathbf{m}$
can be determined via Eq.~(\ref{eq:mass-RC}). Table \ref{tab:schemes}
specifies these sets. In the triplet models there are five
heavy states, but their masses are strongly correlated and a full
on-shell renormalisation would lead to unstable numerical behaviour of
fundamental RCs. Hence we choose to only renormalise a subset of four
heavy masses on-shell, in addition to the light SM-like lepton
masses. The sets  $\bm{\delta}\mathbf{m}$ corresponding to the triplet
models therefore have seven elements, and the choices are explicitly
given in Tab.~\ref{tab:schemes}.

Next, in each model we define a subset of the fundamental RCs in Eq.~\eqref{eq:fundamental-RC} whose real finite parts will be fixed by
the on-shell conditions, while the remaining finite parts are set to 0. We choose the sets
\begin{subequations}\label{setMOS1}
	\begin{align}
		\text{singlet:} &\quad \bm{\delta}\mathbf{M} = \{v \delta y_i, v\delta\lambda, \delta M_E, \delta M_L\}, \qquad \\
		\text{triplet:} &\quad \bm{\delta}\mathbf{M} = \{v \delta y_i, v\delta\lambda, v\delta\lambar, \delta M_E, \delta M_L\}.
	\end{align}
\end{subequations}
with six (seven) elements in the singlet (triplet) models. 
$\overline{\text{MS}}$-renormalisation conditions are imposed on the
remaining fundamental RCs, i.e.\ on $\delta\lambda^L_i,
\delta\lambda^E_i$ (and $\delta\bar{\lambda}$ in the singlet models).

These scheme definitions unambiguously fix the finite parts $\delta^{[0]}\Theta_i$ of all
RCs. In practice, the evaluation of the finite parts of the RCs
$\bm{\delta} \mathbf{M}$ is based on 
Eq.~\eqref{eq:mass-basis-mass-RC} applied to all masses that are
on-shell renormalised. The structure of Eq.~\eqref{eq:mass-basis-mass-RC} leads to
\begin{align}\label{eq:mass-RC-relation-finite}
	\bm{\delta}^{[0]} \mathbf{m} = \Re(\bm{\kappa}) \cdot \Re(\bm{\delta}^{[0]} \mathbf{M}) + \Re(\bm{\delta}^{[0]}\mathbf{u}),
\end{align}
i.e.\ a linear dependence of the (finite part of) on-shell mass
counterterms $	\bm{\delta} \mathbf{m} $ on the fundamental
non-$\overline{\text{MS}}$-renormalised RCs, plus a remainder.
The appropriate coefficients
 $\bm{\kappa}$ are square matrices given in  App.~\ref{app:RC-mat}.
The
(finite part of the) remainder has the structure
$\bm{\delta}^{[0]}\mathbf{u}=\Re(\bm{Y}) \delta^{[0]} v$, where $\bm{Y}\delta v$
corresponds to the respective set of $\delta v$ terms from 
Eq.~\eqref{eq:mass-basis-mass-RC}.
By
inverting the coefficient matrices, the equations can be solved numerically
for the finite parts of the desired fundamental RCs.

\begin{figure}
	\centering
	\begin{subfigure}{.4\textwidth}
		\includegraphics[width=\textwidth]{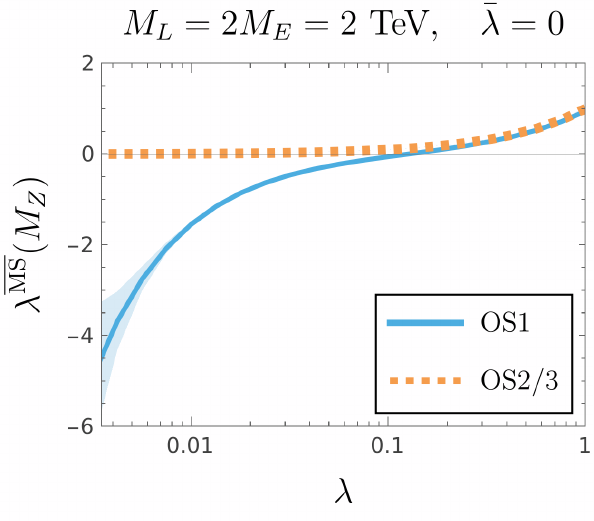}
		\caption{Illustration of a singular limit of OS1.}
		\label{fig:OS1-singular-limits}
	\end{subfigure}\hspace{2cm}
	\begin{subfigure}{.4\textwidth}
		\includegraphics[width=\textwidth]{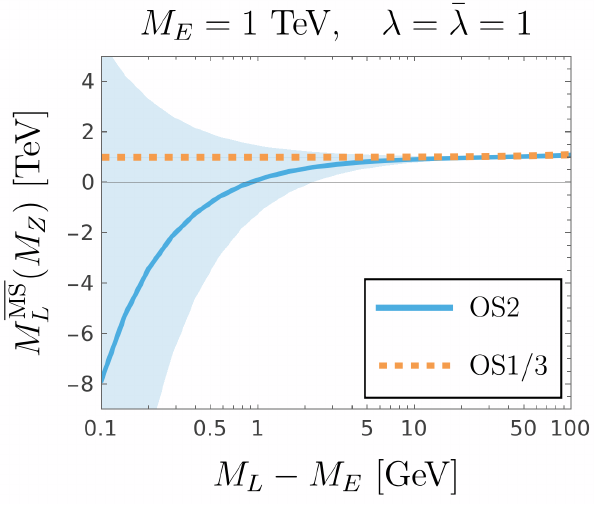}
		\caption{Illustration of a singular limit of OS2.}
		\label{fig:OS2-singular-limits}
	\end{subfigure}
	\vfill
	\caption{Illustration of the \MSbar initial values $\lambda^{\MSbar}(M_Z)$ and $M_L^{\MSbar}(M_Z)$ (including the uncertainty estimates from scale variation discussed in sec.~\ref{sec:theoryuncertainty}) in $L\oplus E$ at the singular limits of OS1 and OS2. 
	In both plots the couplings $\lambda^{L,E}_i$ are set to 0.}
	\label{fig:OS12-singular-limits}
\end{figure}

While convenient for most calculations, this OS1 scheme can lead to
numerical singularities in the limit where the Higgs-induced mixing vanishes 
and the coefficient matrix $\bm{\kappa}$ becomes singular as exemplified in Fig.~\ref{fig:OS1-singular-limits}.

A straightforward way to avoid this issue is to reduce the number of on-shell conditions
imposed on the lepton masses and only determine the set  
\begin{align}
	\bm{\delta}\mathbf{M}  = \{v \delta y_i, \delta M_E, \delta M_L\},
\end{align}
instead of Eq.~\eqref{setMOS1}.
The corresponding choices of on-shell renormalised masses defining
``OS2'' renormalisation schemes are given in the respective column of
Tab.~\ref{tab:schemes}. The corresponding coefficient matrix needed
for solving for the RCs is obtained by dropping columns in $\bm\kappa$
corresponding to $\delta\lambda$ (and $\delta\lambar$ in triplet models) as well as the rows corresponding
to the no-longer on-shell renormalised entries of $\bm{\delta m}$.

While this removes the above singularity, this OS2 scheme instead breaks
down when the Dirac masses are degenerate but the mixing is large, as shown in Fig.~\ref{fig:OS2-singular-limits}. 
To avoid both issues simultaneously, a more refined choice of the masses that are on-shell renormalised must be made.
In particular the two Dirac mass parameters have to be determined from
on-shell conditions on different sectors and it must be ensured that
the corresponding mass counterterms are sensitive
to both $\delta M_E$ and $\delta M_L$. Since this sensitivity depends on the hierarchy between the Dirac mass terms, different regions of parameter space require 
different treatment. We call the resulting third scheme ``OS3'' and
list its definitions in the last column of Tab.~\ref{tab:schemes}. Here it is worth emphasising again that the general strategy
and Eq.~\eqref{eq:mass-RC-relation-finite} is the same for all three schemes and only the choice of $\bm\delta m$ and the resulting coefficient matrix $\bm\kappa$ differ.

\begin{table}[t]
	\centering
	\begin{tabular}{c||Sc|Sc|Sc}
		\textbf{Model} & \textbf{OS1} & \textbf{OS2} & \textbf{OS3} \\ \hline\hline
		$L\oplus E$ & $(\delta m_\nu)_{44}, (\delta m_e)_{44}, (\delta m_e)_{55}$ & $(\delta m_e)_{44}, (\delta m_e)_{55}$ & 
						$(\delta m_\nu)_{44}, \begin{cases}
						 (\delta m_e)_{44} & M_L > M_E\\
						 (\delta m_e)_{55} & M_L \leq M_E
						\end{cases}$ 
		\\ 
		$L\oplus N$ & $(\delta m_\nu)_{44}, (\delta m_\nu)_{55}, (\delta m_e)_{44}$ & $(\delta m_\nu)_{44}, (\delta m_\nu)_{55}$ &
							$(\delta m_e)_{44}, \begin{cases}
							 (\delta m_\nu)_{44} & M_L > M_E\\
							 (\delta m_\nu)_{55} & M_L \leq M_E
						\end{cases}$
		\\ 
		$L_{\frac32}\oplus E$ & $(\delta m_\rho)_{11}, (\delta m_e)_{44}, (\delta m_e)_{55}$ & $(\delta m_e)_{44}, (\delta m_e)_{55}$ &
						$(\delta m_\rho)_{11}, \begin{cases}
							(\delta m_e)_{44} & M_L > M_E\\
							(\delta m_e)_{55} & M_L \leq M_E
						\end{cases}$ 
		\\ 
		$L\oplus E^a$ & $(\delta m_\nu)_{44}, (\delta m_\nu)_{55}, (\delta m_e)_{44}, (\delta m_e)_{55}$ & $(\delta m_e)_{44}, (\delta m_e)_{55}$ &
							$(\delta m_\rho)_{11}, \begin{cases}
							(\delta m_e)_{55} & M_L > M_E\\
							(\delta m_e)_{44} & M_L \leq M_E
						\end{cases}$ 
		\\ 
		$L\oplus N^a$ & $(\delta m_\nu)_{44}, (\delta m_\nu)_{55}, (\delta m_e)_{44}, (\delta m_e)_{55}$ & $(\delta m_e)_{44}, (\delta m_e)_{55}$ &
						$(\delta m_\chi)_{11}, \begin{cases}
							(\delta m_e)_{55} & M_L > M_E\\
							(\delta m_e)_{44} & M_L \leq M_E
						\end{cases}$ 
		\\ 
		$L_{\frac32}\oplus E^a$ & $(\delta m_\rho)_{11}, (\delta m_\rho)_{22}, (\delta m_e)_{44}, (\delta m_e)_{55}$ & $(\delta m_e)_{44}, 
		(\delta m_e)_{55}$ &
						$(\delta m_\nu)_{44}, \begin{cases}
							(\delta m_e)_{55} & M_L > M_E\\
							(\delta m_e)_{44} & M_L \leq M_E
						\end{cases}$ 
	\end{tabular}
	\caption{Choices of the heavy mass counterterms whose real parts are fixed by
the on-shell conditions in each of the schemes. The sets
$\bm{\delta}\mathbf{m}$ are always given by $\Re(\delta m_e)_{11},
\ldots, \Re(\delta m_e)_{33}$ together with the real parts of the entries in
the table. }
	\label{tab:schemes}
\end{table}

\subsection{\texorpdfstring{Redundancy-free renormalisation-group equations and $\overline{\text{MS}}$ counterterms}{Redundancy-free renormalisation-group equations and MS counterterms}}\label{sec:RGE}

Above we have set up three renormalisation schemes for each of the six VLL models
and explained a practical way to determine the finite parts of the
fundamental parameter RCs in Eq. \eqref{eq:fundamental-RC}
from the mass on-shell conditions on the lepton sector Eq.~\eqref{eq:complex-pole-conditiongeneral}.
What is left is to specify the corresponding UV divergent parts such that all renormalised Green functions become finite.

The one-loop UV divergences are the same in all renormalisation
schemes, and are related to the renormalisation-group
$\beta$-functions in the pure $\overline{\text{MS}}$ scheme.
In general, such  $\overline{\text{MS}}$ $\beta$-functions satisfy
\begin{align}\label{eq:general-beta-function}
	\beta(\Theta_i) = -\xi_i \delta^{[1]} \Theta_i + \frac{\partial\delta^{[1]}\Theta_i}{\partial\Theta_j} \xi_j \Theta_j = \sum_\ell 2\ell \; \delta^{[1]}_\ell \Theta_i,
\end{align}
where $\delta^{[1]}_{\ell} \Theta_i$ denotes the $\ell$-loop part of
the $1/\epsilon$ pole of the RC for $\Theta_i$ and $\xi_i=1$ for the Yukawa and gauge couplings, $2$ for the quartic Higgs coupling, and
0 for the mass parameters.

The renormalisation-group running of the \MSbar parameters is given by $\frac{d\Theta_i}{d t} =
\beta(\Theta_i)$ with $t\equiv \ln(Q/Q_0)$, where $Q$ denotes the renormalisation scale and $Q_0$ some reference scale (set to $M_Z$ in the following).  

The \MSbar $\beta$-functions are useful in their own right
in many respects; hence we temporarily switch to the
$\overline{\text{MS}}$ scheme, where all renormalisation constants
consist only of pure $1/\epsilon$ poles\footnote{%
after rescaling the renormalisation scale with the usual factor $\sqrt{e^{\gamma_E}/4\pi}$
} and
discuss here how to determine the relevant
$\beta$-functions up to the two-loop level.
There are generic results for two-loop $\beta$-functions available in
the literature \cite{Machacek:1983tz,Machacek:1983fi,Luo:2002ti,Schienbein:2018fsw,Steudtner:2024teg} (we use the package \texttt{RGBeta}
\cite{Thomsen:2021ncy}). However, specialising these results to VLL
models does not lead to appropriate $\beta$-functions because these
references treat the redundancy between field and parameter
renormalisation differently: they assume hermitian
field renormalisation matrices for gauge-eigenstate fields while
allowing unphysical, off-diagonal Yukawa couplings and Dirac mass
parameters compared to our Eqs.~(\ref{leptonsectorparameters},
\ref{eq:fundamental-RC}). 

To formalise the discussion we denote the $\beta$-functions
corresponding to the generic literature results as $\beta'$ and the
corresponding parameters $\Theta_i'$. The (smaller) set of redundancy-free parameters
corresponding to Eqs.~(\ref{leptonsectorparameters},
\ref{eq:fundamental-RC}) are called $\Theta_i$. It is no problem to
remove the redundancy at tree level, i.e.\ to
assume the specific choice $\Theta_i'=\Theta_i$ at some fixed starting
scale. However, the more restrictive treatment of field
renormalisation then enforces $\delta \Theta_i'\ne\delta\Theta_i$ and
hence different $\beta$-functions.

We illustrate the relationship between the different RCs and $\beta$-functions
explicitly for the $L\oplus E$ model. In this model the redundant RCs
are the ones for the off-diagonal Dirac masses  $\delta M_{L}^i$ and
$\delta M_E^i$ and the off-diagonal SM Yukawa couplings $\delta
y_{ij}$. In the models without either $L$ or $E$ the respective Dirac mass terms
are automatically diagonal, but the structure of the relationship remains the same.
Assuming  $\Theta_i'=\Theta_i$, the redundancy transformation
Eq.~(\ref{eq:redundancy-transformation})
in the $L\oplus E$ model must involve two matrices 
$V_{L/R,0} \in $U(4) of the form
$V_{L/R,0} = \mathbb{1} + \delta V_{L/R}$ to relate the redundant set
of bare parameters $\Theta'_{i,0}$ to the redundancy-free set
$\Theta_{i,0}$. Specifically for $L\oplus E$  we have the relations
\begin{subequations}\label{eq:redundancy-transform-bare}
	\renewcommand\arraystretch{1.2}
	\begin{align}
		\begin{pmatrix}	y_{ij,0}' & \lambda^{E\prime}_{i,0} \\ \lambda^{L\prime}_{j,0} & \lambda'_0 \end{pmatrix} &=
		V_{L,0}^\dagger	\begin{pmatrix}	y_{i,0} \delta_{ij} & \lambda^E_{i,0} \\ \lambda^L_{j,0} & \lambda_0 \end{pmatrix} V_{R,0} ,\\
		\begin{pmatrix}	M_L^{i\prime} \\ M'_L \end{pmatrix}_0 &= V_{L,0}^\dagger \begin{pmatrix} 0 \\ M_L \end{pmatrix}_0, \\
		\begin{pmatrix}	M_E^{i\prime}  & M'_E \end{pmatrix}_0
                &= \begin{pmatrix} 0 & M_E \end{pmatrix}_0 V_{R,0} ,
	\end{align}
\end{subequations}
where  the transformed bare $M_{L,E}^i$ and off-diagonal $y_{ij}$ on the right-hand
side vanish.

Next, we perform the expansion up to two-loop order
and consider only the $1/\epsilon$ pole relevant for the $\beta$-functions
\begin{align}
	\delta^{[1]} V_{L/R} = \delta_1^{[1]} V_{L/R} + \delta_2^{[1]} V_{L/R} + \O(\hbar^3).
\end{align}
The hermitian part at each order is fixed by the unitarity condition $V^\dagger V = \mathbb{1}$, resulting in
\begin{align}
	\delta_1^{[1]} V_{L/R} + \delta_1^{[1]} V_{L/R}^\dagger = 0, \qquad
	\delta_2^{[1]} V_{L/R} + \delta_2^{[1]} V_{L/R}^\dagger = -2 \delta_1^{[1]} V_{L/R}^\dagger \; \delta_1^{[1]} V_{L/R},
\end{align}

while the anti-hermitian part is fixed by  requiring Eq.~(\ref{eq:redundancy-transform-bare}). Introducing the abbreviations
\begin{subequations}
	\begin{align}
		\delta_1^{[1]} \M_L^i &= \delta_1^{[1]} M_L^{i\prime}, \qquad \,\delta_2^{[1]} \M_L^i = \delta_2^{[1]} M_L^{i\prime} + \delta_2^{[1]}\widetilde{M}_L^i, \\
		\delta_1^{[1]} \M_E^i &= \delta_1^{[1]} M_E^{i\prime}, \qquad \delta_2^{[1]} \M_E^i = \delta_2^{[1]} M_E^{i\prime} + \delta_2^{[1]} \widetilde M_E^i, \\[.3cm]
		\delta_1^{[1]} \Upsilon_{ij} &= \delta_1^{[1]} y'_{ij} + \delta_1^{[1]} V_{L,i4}^A\; \lambda^L_j - \lambda^E_i \;\delta_1^{[1]} V_{R,4j}^A, \\
		\delta_2^{[1]} \Upsilon_{ij} &= \delta_2^{[1]} y'_{ij} + \delta_2^{[1]} V_{L,i4}^A\; \lambda^L_j - \lambda^E_i \;\delta_2^{[1]} V_{R,4j}^A + \delta_2^{[1]} \widetilde\Upsilon_{ij},
	\end{align}
\end{subequations}
where the one-loop RC contribution to the two-loop terms is given
\begin{subequations}
	\begin{align}
		\begin{split}
			\delta_2^{[1]} \widetilde{\Upsilon}_{ij} &= \delta_1^{[1]} V_{L,ik} \; \delta_1^{[1]} y'_{kj} \;+\; \delta_1^{[1]} V_{L,i4} \; \delta_1^{[1]} \lambda^{L\prime}_j 
			\;-\; \delta_1^{[1]} y'_{ik} \; \delta_1^{[1]} V_{R,kj}	\;-\; \delta_1^{[1]}\lambda^{E\prime}_i \; \delta_1^{[1]} V_{R,4j} \\
			& \quad + \delta_2^{[1]} V^H_{L,ij} \; y_j + \delta_2^{[1]} V^H_{L,i4} \; \lambda^L_j + y_i \; \delta_2^{[1]}  V^H_{R,ij} + \lambda^E_i \; \delta_2^{[1]} V^H_{R,4j},
		\end{split} \\[.2cm]
		\delta_2^{[1]} \widetilde M_L^i &= \delta_2^{[1]} V^H_{L,i4} \; M_L \;+\; \delta_1^{[1]} V_{L,ik} \; \delta_1^{[1]} M_L^{k\prime} \;+\; \delta_1^{[1]} V_{L,i4} \; \delta_1^{[1]} M'_L,  \\[.2cm]
		\delta_2^{[1]} \widetilde M_E^i &= \delta_2^{[1]} V^H_{R,4i} \; M_E \;-\; \delta_1^{[1]} V_{R,ki} \; \delta_1^{[1]} M_E^{k\prime} \;-\; \delta_1^{[1]} V_{R,4i} \; \delta_1^{[1]} M'_E,
	\end{align}
\end{subequations}
these conditions yield the following solutions\footnote{Note that the higher order poles, like e.g. $1/\epsilon^2$, can be obtained iteratively from the 
	consistency condition
	\begin{align}
		\frac{\partial \delta^{[n-1]}\Theta_i}{\partial\Theta_j} \beta(\Theta_j) = -\xi_i \delta^{[n]}\Theta_i +  \frac{\partial \delta^{[n]}\Theta_i}{\partial\Theta_j} \xi_j\Theta_j
	\end{align}
	where the right hand side has the same loop-order decomposition as in Eq.~\eqref{eq:general-beta-function}, but they are not needed for the purposes of this paper.
	
}
\begin{alignat}{3}
	\delta_\ell^{[1]} V^A_{L,i4} &= - \frac{\delta_\ell^{[1]} \M_L^i}{M_L}, \qquad &&
	\delta_\ell^{[1]} V^A_{L,ij} &&=  \frac{\delta_\ell^{[1]}\Upsilon_{ji}^* y_i + \delta_\ell^{[1]}\Upsilon_{ij} y_j^*}{|y_i|^2 - |y_j|^2}, \\
	\delta_\ell^{[1]} V^A_{R,4i} &= - \frac{\delta_\ell^{[1]} \M_E^i}{M_E}, \qquad &&
	\delta_\ell^{[1]} V^A_{R,ij} &&=  \frac{\delta_\ell^{[1]}\Upsilon_{ji}^* y_j + \delta_\ell^{[1]}\Upsilon_{ij} y_i^*}{|y_i|^2 - |y_j|^2},
\end{alignat}
Using these two-loop results for the matrices $V_{L/R,0}$ in
Eq.~\eqref{eq:redundancy-transform-bare}, allows translating bare
redundant parameters $\Theta_{i,0}'$ into bare non-redundant
parameters $\Theta_{i,0}$. This already provides us with the 
$1/\epsilon$ poles of the required one-loop RCs in
Eq.~(\ref{eq:fundamental-RC}). 
Via Eq.~\eqref{eq:general-beta-function}, an equivalent translation
yields the two-loop $\beta$-functions for the redundancy-free
parameters. 

Finally, the field redefinition Eq.~\eqref{eq:redundancy-transformation} leads to an anti-hermitian shift of the field renormalisation
constants that has to be included in order to keep all (off-shell) Green functions finite. For the mass-eigenstate field RCs this shift
implies at one-loop
\begin{align}\label{eq:anti-hermitian-field-RC-shift}
	\big(\delta^{[1]}_1 Z^{f,A}_{L,R}\big)_{ab} = 2 \, \big(U^{f\dagger}_{L,R}\big)_{a\tilde{c}} 
	\big(\delta_1^{[1]} V_{L,R}\big)_{\tilde{c}\tilde{d}} \big(U_{L,R}^f\big)_{\tilde{d}b}.
\end{align}
Note that the off-diagonal part is also determined by the on-shell conditions (see Eq.~\eqref{eq:OS-solution}) in terms
of the mass RCs and automatically agrees with the above expression when the redundancy free RCs are inserted. However,
notably the OS conditions did not determine the diagonal part which are now fixed by Eq.~\eqref{eq:anti-hermitian-field-RC-shift}.
It is also worth pointing out that this diagonal part is only non-zero in the case of CP violation.

%% file: Potential.tex
\section{The effective potential and vacuum stability}\label{sec:Potential}

The OS renormalisation schemes discussed above allow us to compute next-to-leading order (NLO) corrections to observables in the VLL models
as a function of the BSM parameters while simultaneously ensuring
that the predicted SM lepton masses match the experiment exactly.
Here we focus on the implications of VLLs for Higgs vacuum stability.
Technically, we use the ``OS3'' scheme to derive the \MSbar values
of the coupling parameters at the EW scale. At one-loop order the relation between the two schemes is given by\footnote{Note that in our OS schemes all of the lepton Yukawa
	couplings have a residual renormalisation scale dependence due to the mixed OS/\MSbar conditions. On the other hand, the gauge, Higgs potential and quark Yukawa couplings 
	are fixed entirely on-shell and thus do not depend on the renormalisation scale.}
\begin{align}\label{eq:OS-to-MS}
	\Theta_i^\text{\MSbar} = \Theta_i^\text{OS} + \delta^{[0]} \Theta_i + \O(\hbar^2)
\end{align}
where $\delta^{[0]} \Theta_i$ denotes the finite part of the OS RCs.
We then solve the redundancy-free
two-loop renormalisation group equations (RGEs) of Sec.~\ref{sec:RGE} to obtain the couplings at
the high scales, with which we compute the 
effective Higgs potential at high field values.
In this way, one-loop fixed-order contributions are taken into account
and large logarithms are resummed at the next-to-leading log (NLL) level.

\subsection{Effective potential and instability scale}
The RG-improved effective potential is given by
\begin{align}
	V_\text{eff}(h) = -\mu^2(t)\frac{h^2}{2}  + \lambda_\Phi(t)\frac{h^4}{4} + \Delta V_\text{eff}(h),
\end{align}
where $\mu^2$, $\lambda_\Phi$ and the parameters in $\Delta V_\text{eff}(h)$ are the running \MSbar couplings evaluated at $Q=h$.
The radiative corrections $\Delta V_\text{eff}$ are given by the usual
Coleman-Weinberg (CW) formula \cite{Coleman:1973jx}.
While the quark, gauge and Goldstone boson contributions are
functionally the same as in the SM (cf. Refs.~\cite{Sher:1988mj,Sher:1993mf,Degrassi:2012ry,Hiller:2024zjp}) and depend on
the VLL couplings only through the loop corrections to the initial values and RGEs,
the lepton contributions are modified by the VLL and are given by
\begin{align}\label{eq:CW-leptons}
	\Delta_\text{lep}^{1\ell} V_\text{eff}(h) = \sum_{f=\nu,e,\rho} \sum_{a=1}^{n_f}\frac{m_{f_a}^4(h)}{16\pi^2} \bigg[ \ln\bigg(\frac{Q^2}{m_{f_a}^2(h)}\bigg) + \frac{3}{2}\bigg],
\end{align}
where the (diagonal) field-dependent mass matrix is given by
\begin{align}
	m_f(h) = U_L^{f\dagger}(h) \bigg[\frac{h}{\sqrt{2}} \mathcal{Y}^h_f(Q) + \mathcal{M}_f^D(Q)\bigg] U_R^f(h).
\end{align}
Here $\mathcal{Y}^h_f$ and $\mathcal{M}_f^D$ denote the running ($\overline{\text{MS}}$) Yukawa and Dirac mass matrices and $U_{L,R}(h)$ 
diagonalise the field-dependent mass matrix.

The Goldstone boson contributions introduce both a gauge dependence as well as an imaginary part, since their field-dependent masses typically become negative.
However, we have checked that overall their contributions are small compared to the VLL and top-quark one-loop corrections, in particular at large field values,
and can therefore be ignored for the purposes of this paper, avoiding these issues.

At low values $h\simeq \sqrt2 v$, the Dirac masses dominate in Eq.~\eqref{eq:CW-leptons}, leading to a large contribution to the Higgs mass $\propto M_{L,E}^2$ that 
is in principle compensated by the OS conditions on $\delta \mu^2$. However,
if the translation of $\mu^2$ from OS to \MSbar is performed at one-loop (using Eq.~\eqref{eq:OS-to-MS}) 
while the CW corrections to the effective potential are computed in terms of the \MSbar parameters, there is a residual 2-loop mismatch that is still proportional
to $M_{L,E}^2$. This artifact of the scheme conversion can be large and lead to an incorrect shift of the EW minimum away from $h=\sqrt{2}v$.
To account for this effect, instead of using Eq.~\eqref{eq:potential-parameter-RC} and \eqref{eq:OS-to-MS} to determine the \MSbar value of $\mu^2$, we derive it from the
minimisation condition on the effective potential, similar to the procedure common in spectrum generators, see e.g.\ \cite{Porod:2011nf,Athron:2017fvs},
\begin{align}
	\frac{d V_\text{eff}}{dh}\Big|_{h=\sqrt{2}v} = 0.
\end{align}
For the lepton contribution the derivative is given by
\begin{align}
	\frac{d}{dh} \Delta_\text{lep}^{1\ell} V_\text{eff}(h) = \sum_{f=\nu,e,\rho} \sum_{a=1}^{n_f}\frac{m_{f_a}^3(h)}{4\pi^2} \bigg[ \ln\bigg(\frac{Q^2}{m_{f_a}^2(h)}\bigg) + 1\bigg] \frac{d m_{f_a}}{dh},
\end{align}
where the derivative of the lepton masses is given by
\begin{align}
	\frac{d m_{f_a}}{dh} = \frac{1}{\sqrt2}\Re\bigg[ U_L^\dagger(h)  \mathcal{Y}^h_f(Q) U_R(h)\bigg]_{aa}.
\end{align}

Like in the SM, the loop-corrected effective potential may develop
deeper minima than the EW one at high field values $h$,
signalling an instability of the EW vacuum. To quantitatively describe
the instability scale we first define  an effective quartic Higgs coupling as 
\begin{align}
	\lambda_\Phi^\text{eff}(h) = \frac{4}{h^4}\Big[V_\text{eff}(h) - V_\text{eff}(h_\text{min}^\text{EW})\Big],
\end{align}
where $h_\text{min}^\text{EW}\simeq \sqrt{2} v$ is the field value at the minimum of
the RG-improved effective potential near the electroweak scale.
We then define the instability scale $\Lambda$ as the smallest value of the
field $h$ where a deeper minimum starts to develop, i.e.\ where
$\lambda_\Phi^{\text{eff}}(\Lambda)= 0$.
We also consider the scale $\bar\Lambda$ where the running quartic
coupling vanishes, $\lambda_\Phi(\bar\Lambda) = 0$, as a common proxy for
$\Lambda$.  

The plots in Fig.~\ref{fig:lambdaEffplots} illustrate various
possibilities for the behaviour of the effective coupling
$\lambda_\Phi^\text{eff}(h)$. For each curve, a specific VLL model (either
$L\oplus E$ or  $L\oplus E^a$) and specific
parameter values (either only $\bar\lambda\ne0$ or
$\bar\lambda=\lambda=\lambda_2^{L,E}\ne0$ with values as displayed in
the plots) were chosen. In a wide parameter region, the behaviour is
as the one of the red/orange lines in
Fig.~\ref{fig:UsualBehaviour-Illustration}: the effective potential
and both $\lambda_\Phi^\text{eff}(h)$ and $\lambda_\Phi(h)$ decrease
at large field values, leading to a sign change at a certain scale (in
the plots, the sign change happens around $h\sim10^{9\ldots11}$
GeV). In these cases, the position of the sign change of
$\lambda_\Phi^\text{eff}(h)$ is the 
instability scale $\Lambda$. The dependence of $\Lambda$ on the
fundamental model parameters is continuous in that region.

The dark blue lines in Fig.~\ref{fig:UsualBehaviour-Illustration}
illustrate an exceptional case, a discontinuous transition that happens in the
$L\oplus E^a$ model. Here, for sufficiently small couplings the
effective potential is stabilised and $\lambda_\Phi^\text{eff}(h)$ stays
positive, while for larger couplings $\lambda_\Phi^\text{eff}(h)$
develops two zeros and the instability scale $\Lambda$
discontinuously jumps to a finite value.

At larger values of the BSM parameters, more nontrivial behaviour can
occur as exemplified in Fig.~\ref{fig:PhaseChange-Illustration}. The
complex interplay of model-dependent Yukawa and gauge structure can
lead to stabilising or destabilising effects, leading to numerous
inflection points and, depending on the exact parameter
choice, multiple zero crossings as seen in orange in
Fig.~\ref{fig:PhaseChange-Illustration} or the disappearance of a
second minimum as seen in blue. Both cases lead to
discontinuous jumps in the instability scale as a function of the VLL couplings $\lambda_i$.
Our numerical implementation ensures that the correct
  instability scale $\Lambda$ is found also in such special cases.

\subsection{Estimate of the instability scale uncertainty}\label{sec:theoryuncertainty}

\begin{figure}[t]
	\centering
	\begin{subfigure}[t]{.49\textwidth}
		\includegraphics[width=.8\textwidth]{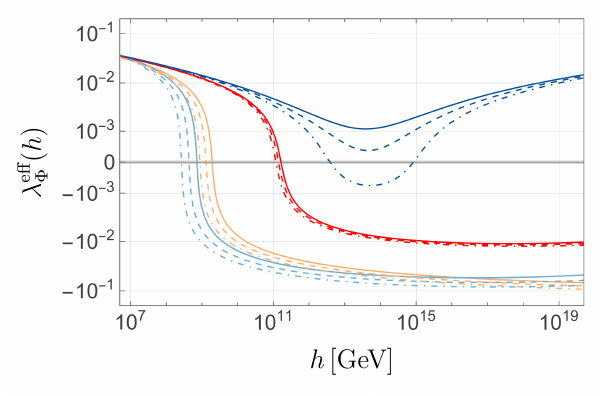}
		\raisebox{.75cm}{\includegraphics[width=.15\textwidth]{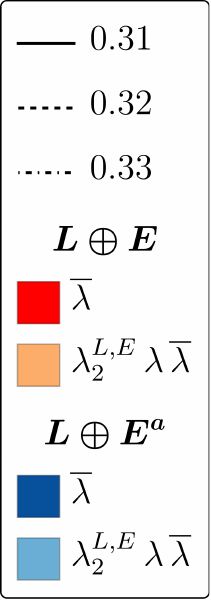}}
		\caption{Comparison of the typical behaviour of the singlet and triplet models for one or four active VLL couplings
			at moderate VLL couplings. }
		\label{fig:UsualBehaviour-Illustration}
	\end{subfigure}
	\begin{subfigure}[t]{.49\textwidth}
		\includegraphics[width=.8\textwidth]{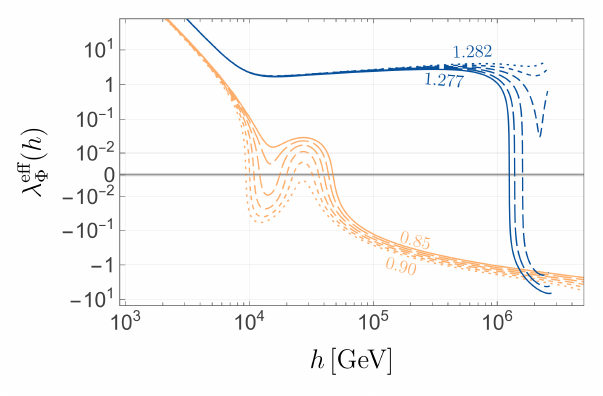}
		\raisebox{1.5cm}{\includegraphics[width=.15\textwidth]{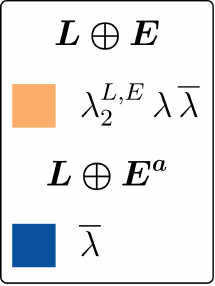}}
		\caption{Illustration of the development of discontinuous jumps in $\Lambda$. The VLL couplings are varied uniformly between the indicated values. } 
		\label{fig:PhaseChange-Illustration}
	\end{subfigure}
\caption{Plots of $\lambda_\Phi^\text{eff}(h)$ around the instability scale $\Lambda$ for different parameter points (at $M_L = M_E = \SI{10}{TeV}$) in the $L\oplus E$ and $L\oplus E^a$ model. 
	\textbf{(a)} shows the typical behaviour for moderate couplings, while \text{(b)} illustrates the formation of discontinuous jumps in the instability scale at large couplings.}
\label{fig:lambdaEffplots}
\end{figure}

Although the effective potential is computed using two-loop RGEs and
one-loop EW-scale matching, a theory uncertainty remains. Clearly,
given the qualitatively diverse possibilities visible in
Fig.~\ref{fig:lambdaEffplots}, this uncertainty depends on the model
and the chosen input parameter values. Here we
provide a conservative estimate for this uncertainty that can be
numerically evaluated at each model parameter point. 

There are three main sources of uncertainty in determining the
effective potential, $\lambda_\Phi^\text{eff}$ and ultimately
$\Lambda$. The first is parametric, driven mainly by the measurement
uncertainty  
of $M_h = 125.13(11)$ GeV, $m_t = 172.60(27)$ GeV and
$\alpha_s = 0.1180(9)$. We estimate the parametric uncertainty of
$\Lambda$ conservatively as the envelope of the $1\sigma$ variations
of all three input parameters. The relative parametric uncertainty is thus denoted as
\begin{align}
	\Delta_\text{par}\Lambda = \frac{\Lambda_{\sigma=\pm1}}{\Lambda}.
\end{align}

A second source of theory uncertainties stems from the missing higher-order corrections when converting between the OS and \MSbar values of the coupling constants in Eq.~\eqref{eq:OS-to-MS}.
In addition to the complete one-loop corrections we also include the two-loop QCD and $y_t$ contributions to $\lambda_\Phi$, $y_t$ and $g_s$.
The remaining conversion uncertainty is estimated via scale variation where the central value $Q_0=M_Z$ is varied between $Q=M_Z/2... 2M_Z$.
This simulates missing two-loop VLL and three-loop QCD contributions and yields the (asymmetric) uncertainty\footnote{Here we do not take into account the residual scale dependence
of the BSM Yukawa couplings in our on-shell scheme(s). This effect contributes to $\lambda_\Phi(t)$ at the same (two-loop) level as the RG running
and thus does not change the structure of the uncertainty estimate.}
\begin{align}
	\Delta_Q \Lambda = \{\Lambda_{Q_0=\frac{M_Z}{2}}, \Lambda_{Q_0=2M_Z}\} / \Lambda
\end{align}

A third source of uncertainty stems from the truncation of the RGE. As discussed above, we take into account the full two-loop
$\beta$-functions including all couplings of the model, resulting in full
NLL accuracy. For large VLL couplings, $\lambda_\Phi$ typically falls off quickly,
such that the uncertainty induced by the running becomes small. On the other hand, for small VLL couplings, the RGE uncertainty is again driven by
the QCD corrections. This uncertainty can be determined efficiently by linearising the system in the truncation error
\begin{align}
	\frac{d}{dt} \bm{\Delta\Theta} = \frac{\partial\beta(\bm{\Theta})}{\partial\bm{\Theta}} \bm{\Delta\Theta} + \Delta_3\beta(\bm{\Theta}),
	\qquad\text{for}\qquad
	\bm{\Theta} = \{\lambda_\Phi, y_t, g_s \}.
\end{align}
Here, $\Delta_3\beta(\bm{\Theta})$ denotes the 3-loop contribution to
the $\beta$-functions purely governed by QCD and $y_t$.
The resulting uncertainty on the instability scale is
\begin{align}
	\Delta_\text{RGE} \Lambda \approx \Delta_\text{RGE} \bar\Lambda = \exp(- \frac{\Delta\lambda_\Phi(\bar\Lambda)}{\beta(\bar\Lambda)}).
\end{align}
This uncertainty, however, is typically smaller than the parametric and initial-scale uncertainties, and can mostly\footnote{We have found, in contrast, that in a leading-log analyses the RGE truncation error is similar to the other two sources of uncertainty.} be neglected.

Lastly, we need to address how to combine the above estimates to obtain the total uncertainty of the
instability scale. In principle the three sources are independent, however, since the instability scale
depends discontinuously on the parameters (see discussion above) a simple Gaussian combination typically 
drastically overestimates the uncertainty close to the non-analytic jumps.
A simple fix is to take the pointwise maximum of the three uncertainties as
an estimate of the overall uncertainty
\begin{align}
	\log_{10}\Delta\Lambda = \max\Big\{|\log_{10}\Delta_\text{par}\Lambda|,~|\log_{10}\Delta_Q\Lambda|,~|\log_{10}\Delta_\text{RGE}\Lambda|\Big\},
\end{align}
in positive and negative directions each. Testing has shown that in
continuous regions this uncertainty estimate is consistent with
standard uncertainty propagation as here mostly only one source of
uncertainty dominates, while in discontinuous regions the maximum
avoids double counting of uncertainties that would be present when
e.g.\ summing them in quadrature.

\begin{figure}[t]
	\centering
	\includegraphics[width=.5\textwidth]{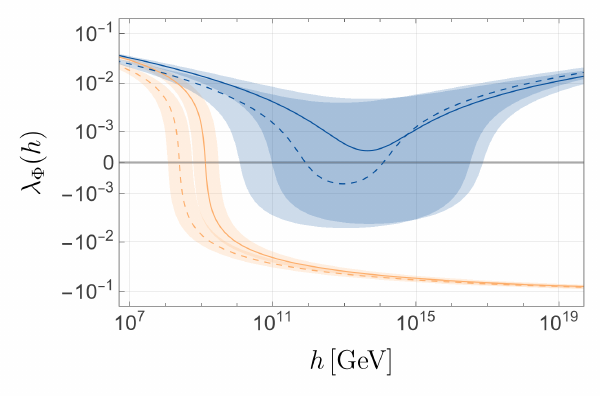}
	\raisebox{1.5cm}{\includegraphics[width=.08\textwidth]{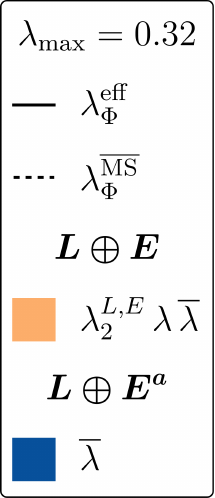}}
		\caption{Comparison of the effective quartic Higgs coupling $\lambda_\Phi^\text{eff}(h)$ (solid)
			and the running \MSbar coupling $\lambda_\Phi^{\overline{\text{MS}}}$ (dashed) in $L\oplus E$ (with 4 active Yukawa couplings)
			and $L\oplus E^a$ (with one active Yukawa coupling). The parameter points correspond to those already shown in Fig.~\ref{fig:UsualBehaviour-Illustration}
			for the indicated VLL couplings set to $0.32$. We also show the uncertainty estimate described in the text.}
		\label{fig:lamHvEffPot-Illustration}
\end{figure}

The uncertainty bands for the (effective) couplings
$\lambda_\Phi^\text{eff}(h)$ and $\lambda_\Phi(h)$
are derived analogously from
the pointwise maximum of the parametric and initial-scale
uncertainties, but neglecting the RGE truncation uncertainty.
Fig.~\ref{fig:lamHvEffPot-Illustration} shows these uncertainty bands and a comparison between the effective quartic Higgs coupling $\lambda_\Phi^\text{eff}(h)$
and the running \MSbar value $\lambda_\Phi(h)$.
The parameter choices are the same as the ones in
Fig.~\ref{fig:UsualBehaviour-Illustration} and correspond to the generic situation of a singlet model and the situation close to a discontinuous transition in a triplet model.
The zeros of the curves correspond to the
instability scale $\Lambda$ and its proxy $\bar\Lambda$. 
Generally, the proxy is smaller than $\Lambda$ because of the additional positive
terms in the effective potential contained in $\lambda_\Phi^\text{eff}$ but not in the running coupling $\lambda_\Phi$. At large $h$ these terms are
suppressed and the two couplings approach each other. Overall,
$\bar\Lambda$  is a good approximation for $\Lambda$ in most regions
of the parameter space and gives 
a more conservative estimate of the instability scale than the
effective quartic Higgs coupling.
The uncertainties displayed in
Fig.~\ref{fig:lamHvEffPot-Illustration} are small in large parts of
the parameter space and become larger only in regions close to the discontinuous
transition. This translates into a significant uncertainty on the
location of the zero, i.e.\ on the instability scale $\Lambda$, close
to the transition. However, in other parts of the parameter space 
the determination of the instability scale turns out to be robust.

\subsection{Bounds from vacuum stability}

Here we apply the loop-corrected and RGE-improved effective potential
to determine the vacuum instability scale $\Lambda$ as a function of
the VLL model parameter space. As an overview, in the limit where the VLL
couplings vanish our computation has largely two outcomes depending on the models.
In singlet models, the result becomes similar to the well-known SM instability scale of $\Lambda \sim \SI{10^{11}}{GeV}$ within the uncertainty bands.
In triplet models, the sizeable additional gauge contributions stabilise the potential at large field values and we find that the instability scale is pushed beyond the Planck scale to $\Lambda \sim \SI{10^{19}}{GeV}$.
For nonzero VLL couplings the vacuum tends to become more unstable in all models, and in large parts of the parameter spaces the VLL interactions lead
to a lower instability scale than in the SM. 
Here we do not impose the requirement
of a fixed minimum instability scale as a criterion for the validity
of the models, but different proposals for such a requirement have
been used in the literature, ranging from Planck safety
\cite{Hiller:2024zjp}, i.e.\ requiring $\Lambda > M_\text{Pl} \approx
\SI{10^{19}}{GeV}$, to smaller bounds of $\Lambda > \SI{10^{5}}{GeV}$
\cite{Adhikary:2024esf}, or requiring the minimum instability scale to
be sufficiently above the physical BSM masses.

\subsubsection{Interplay with EWPOs}

Constraints from vacuum stability are naturally complementary to
constraints from electroweak precision observables (EWPOs). Neither kind
of constraint depends on flavour violation but generically
limit the possible values of BSM couplings to the electroweak sector.
In the literature, EWPO constraints on VLL models have been
investigated, and the constraints essentially limit the combinations
$\lambda_i^{L,E}v/M_{L,E}$ between VLL Yukawa couplings and Dirac masses
\cite{Kannike:2011ng,Crivellin:2020ebi,Daberstiel:2026wxe}.
Vacuum stability now allows us to
constrain the set of VLL Yukawa couplings directly, either
independently of the VLL masses or with different scaling relations than
in EWPO observables.

\begin{figure}[t]
	\centering
	\begin{subfigure}[t]{\textwidth}
		\centering
		\sbox{\figbox}{\includegraphics[width=.3\textwidth,trim={0 1 .75 0},clip=true]{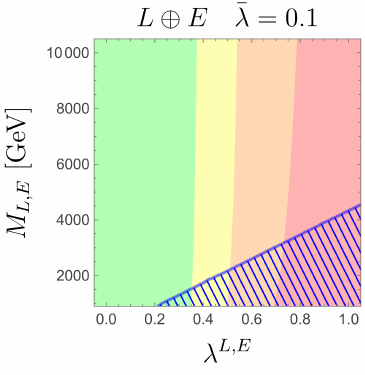}}
		\usebox{\figbox}
		\hfill
		\includegraphics[height=\ht\figbox,trim={22.5 0 .75 0},clip=true]{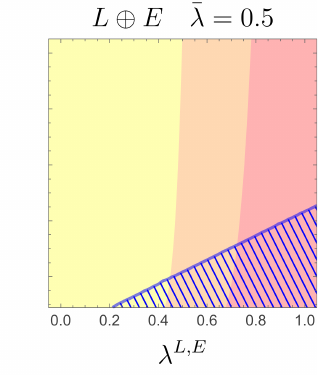}
		\hfill
		\includegraphics[height=\ht\figbox,trim={22.5 0 .75 0},clip=true]{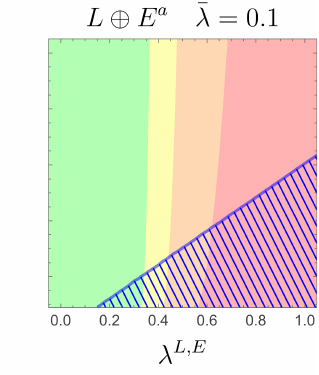}
		\hfill
		\includegraphics[height=\ht\figbox,trim={22.5 0 0 0},clip=true]{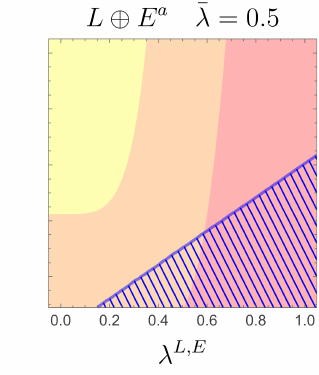}
		\caption{Contours of $\Lambda$ for $\lambda^{L}_2=\lambda^E_2$ and $M_L=M_E$.}
		\label{fig:LLR-M12-EWPO}
	\end{subfigure}\vspace*{1em}
	\begin{subfigure}[t]{\textwidth}
		\centering
		\sbox{\figbox}{\includegraphics[width=.275\textwidth,trim={0 1 1 0},clip=true]{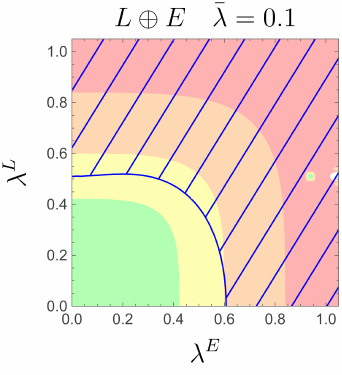}}
		\usebox{\figbox}
		\hfill
		\includegraphics[height=\ht\figbox,trim={5 0 0 0},clip=true]{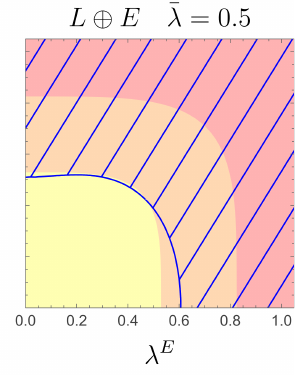}
		\hfill
		\includegraphics[height=\ht\figbox,trim={5 0 0 0},clip=true]{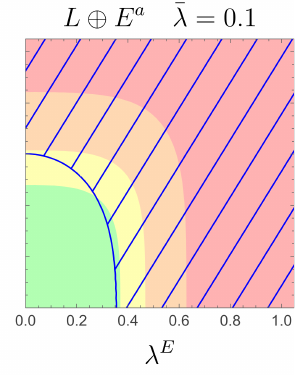}
		\hfill
		\includegraphics[height=\ht\figbox,trim={5 0 0 0},clip=true]{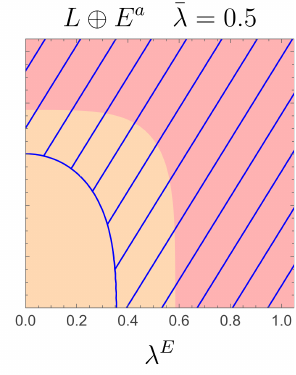}
		\caption{Contours of $\Lambda$ as a function of the VLL couplings $\lambda^{L,E}_2$. The Dirac masses are set to $M_{L,E}=2$ TeV.}
		\label{fig:LL-LR-EWPO}
	\end{subfigure}
	\caption{Contours of the vacuum instability scale $\Lambda$ and the EWPO $2\sigma$ exclusion limits (blue hatched) for the $L \oplus E$ and $L \oplus E^a$ models in the $\lambda^{L,E}_2$--$M_{L,E}$ plane (top)
		 and $\lambda^{L}_2$--$\lambda^E_2$ plane (bottom). 
		 The colours correspond to values of $\Lambda > 10^{10}$ GeV (green), $10^8$ -- $10^{10}$ GeV (yellow), $10^6$ -- $10^8$ GeV (orange) and $\Lambda<10^6$ GeV (red). All VLL Yukawa couplings not shown are set to 0.}
	\label{fig:EWPOvsVacstab}
\end{figure}

We illustrate this complementarity in
Fig.~\ref{fig:EWPOvsVacstab} for the $L \oplus E$ and $L \oplus E^a$ model in different slices of the parameter
space. Figure~\ref{fig:LLR-M12-EWPO} shows
the general mass dependence and $\lambda^{L,E}$ dependence of
$\Lambda$. In line with the previous Fig.~\ref{fig:lambdaEffplots}, it
shows that the vacuum gets more stable with 
smaller Yukawa couplings and larger VLL masses. The mass dependence of
$\Lambda$ gets weaker the larger the VLL masses are, leading to
almost vertical contours at  multi-TeV masses, while the EWPO constraints
shown in blue constrain ratios of Yukawa and mass parameters, leading
to triangular exclusion contours. 

In Fig.~\ref{fig:LL-LR-EWPO} the EWPO and vacuum stability constraints
are shown as functions of VLL couplings $\lambda_2^{L}$ vs. $\lambda_2^E$ for fixed mass parameters and $\lambar$. For the chosen mass scale of $2$~TeV, the
parameter regions allowed by EWPO lead to vacuum instability scales of
$\Lambda \gtrsim \SI{10^{6}}{GeV}$. Since the EWPO constraints are inversely
proportional to the VLL masses, increasing the VLL masses will lead to
a relaxation of the EWPO constraints, while the vacuum stability
constraints are only weakly dependent on the VLL masses, amplifying
their importance in comparison with EWPO and surpassing them at larger
masses, depending on the chosen exclusion scale. 
While not shown here, similar results are obtained for other choices of singlet and triplet models and the instability scale contours only change slightly. The only significant differences between models are the EWPO exclusion contours, which are already discussed in detail in \cite{Daberstiel:2026wxe} and are not the focus of this work.

\subsubsection{General Exclusion based on the largest Yukawa coupling}

The instability scale $\Lambda$ is primarily determined by the magnitude of the VLL-Yukawa couplings $\lambdamax = \max_i |\lambda_i|$. The detailed dependence is governed by the number of couplings at or near this magnitude and the singlet or triplet structure of the studied model. 
To analyse the instability scale $\Lambda$ as a function of $\lambdamax$, we look at slices of parameter space where different numbers of Yukawa
couplings are switched on, while the VLL mass scale is
fixed. Equivalently, such an analysis allows us to read off the maximally
acceptable VLL Yukawa coupling $\lambdamax$ under broad conditions,
once a minimum $\Lambda$ is imposed, see also
Ref.~\cite{Adhikary:2024esf} for a similar analysis.

\begin{figure}[t]
	\centering
	\newcommand{\lmaxplotheight}{.1\textheight}
	\begin{subfigure}{\textwidth}
		\sbox{\figbox}{\includegraphics[width=.26\textwidth,trim={0 30 0 0},clip=true]{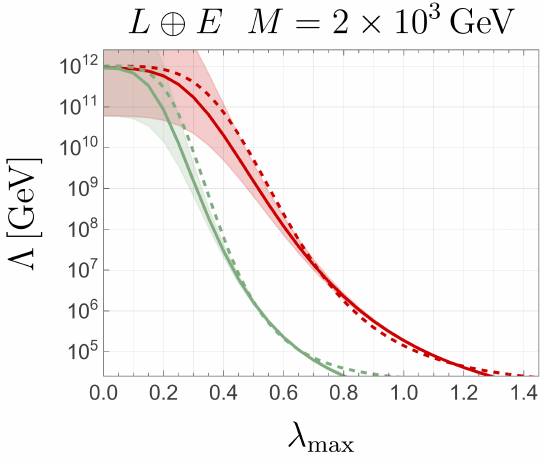}}
		\usebox{\figbox}
		\includegraphics[height=\ht\figbox,trim={25 30 0 0},clip=true]{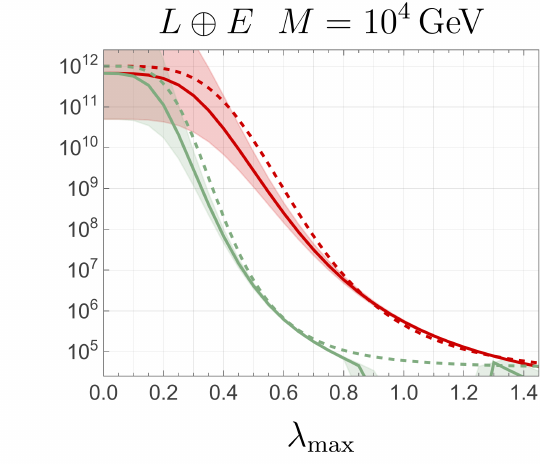}
		\includegraphics[height=\ht\figbox,trim={25 30 0 0},clip=true]{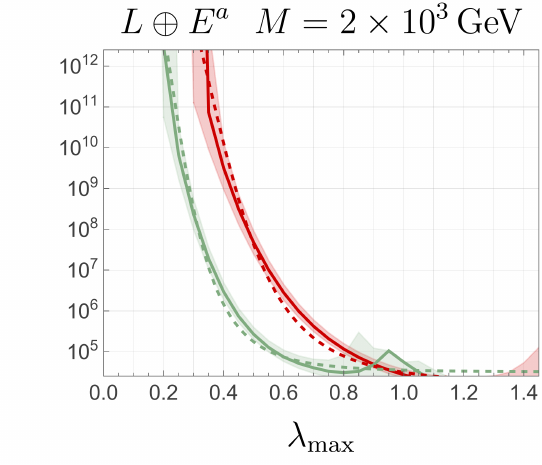}
		\includegraphics[height=\ht\figbox,trim={25 30 0 0},clip=true]{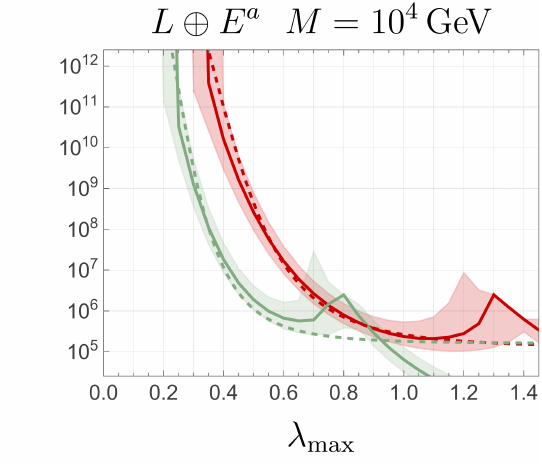}
		\vspace{2pt}
		\\
		\sbox{\figbox}{\includegraphics[width=.26\textwidth,trim={0 30 0 0},clip=true]{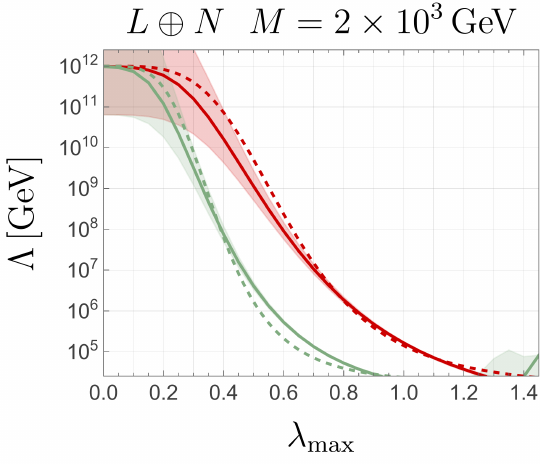}}
		\usebox{\figbox}
		\includegraphics[height=\ht\figbox,trim={25 30 0 0},clip=true]{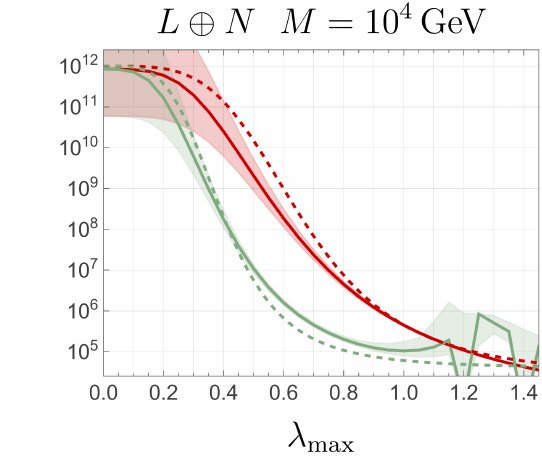}
		\includegraphics[height=\ht\figbox,trim={25 30 0 0},clip=true]{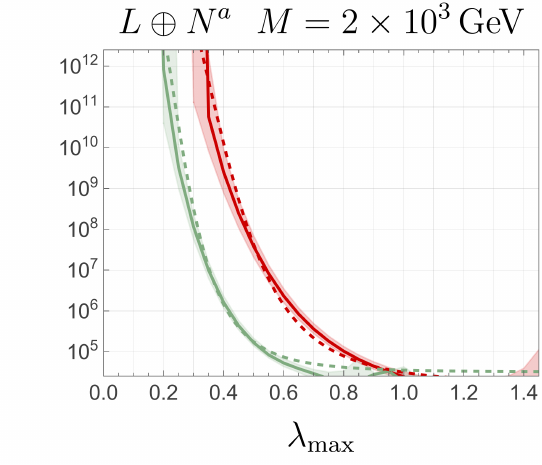}
		\includegraphics[height=\ht\figbox,trim={25 30 0 0},clip=true]{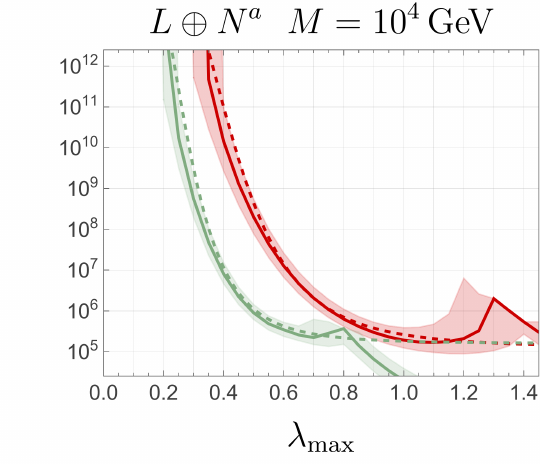}
		\vspace{2pt}
		\\
		\sbox{\figbox}{\includegraphics[width=.26\textwidth,trim={0 0 0 0},clip=true]{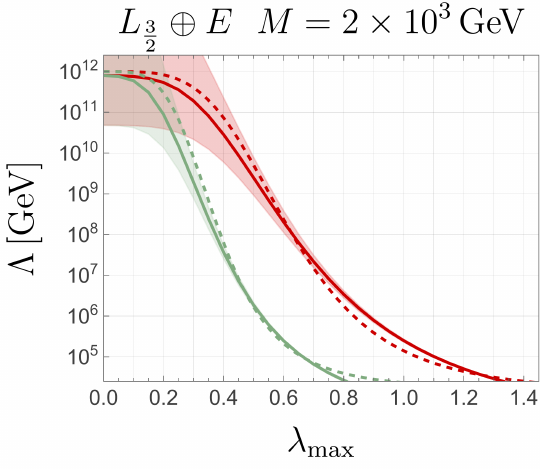}}
		\usebox{\figbox}
		\includegraphics[height=\ht\figbox,trim={25 0 0 0},clip=true]{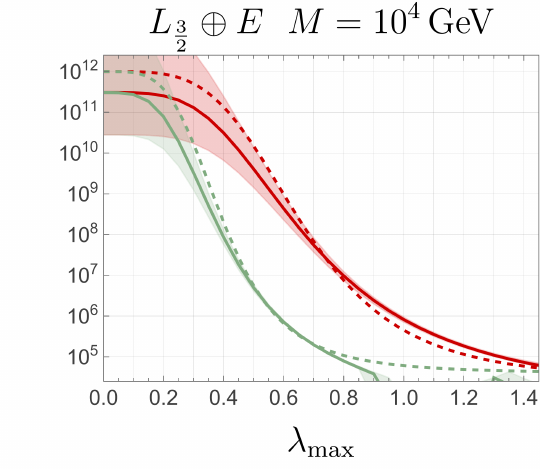}
		\includegraphics[height=\ht\figbox,trim={25 0 0 0},clip=true]{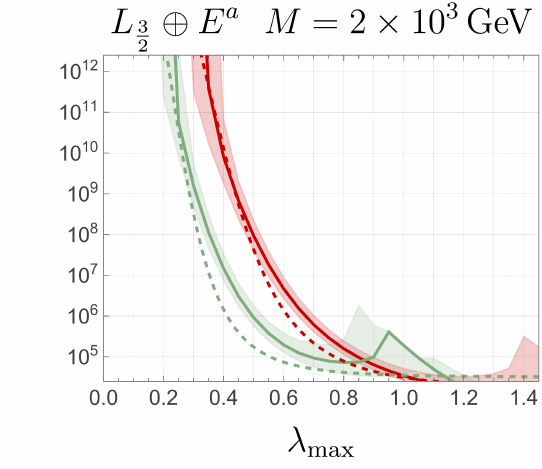}
		\includegraphics[height=\ht\figbox,trim={25 0 0 0},clip=true]{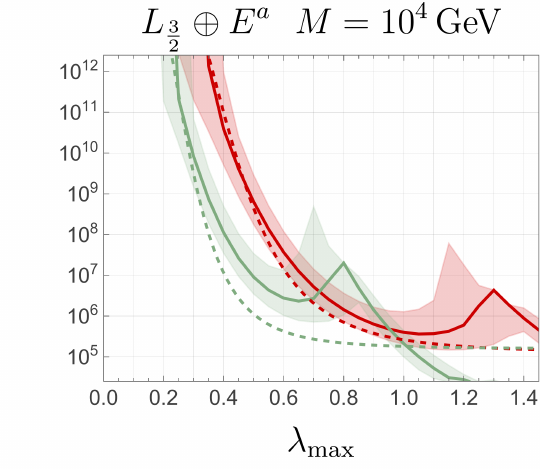}
	\end{subfigure}
	\begin{subfigure}{\textwidth}
		\centering
		\raisebox{-.2cm}{\includegraphics[width=.45\textwidth]{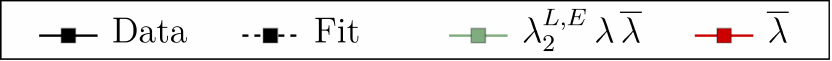}}
	\end{subfigure}
	\caption{Plots of $\Lambda(\lambdamax)$ for the six VLL models. We show the numerical results (solid) with the uncertainty estimate (shaded) 
		and the approximate fit (dashed) for 1 (red) and 4 (green) active Yukawa couplings (as indicated) and $M_L = M_E \equiv M$ at 2 and 10 TeV.}
	\label{fig:Lambdaresult}
\end{figure}

For all six model classes we have performed a scan over a search space
consisting of either one ($\lambdamax = \lambar$) or four
($\lambdamax = \lambar = \lambda = \lambda_2^{L,E}$) nonzero Yukawa
couplings\footnote{It was checked that other choices do not change the
analysis within the uncertainty bands.} and masses of either
$M_L = M_E = \SI{2}{TeV}$ or $\SI{10}{TeV}$. 
The results for all these 24 cases are shown in
Fig.~\ref{fig:Lambdaresult} as the solid lines. The  bands
correspond to the uncertainty estimate as described in
Sec.~\ref{sec:theoryuncertainty}. 

The general trends are similar for all models with VLL singlets (the first
two columns in Fig.~\ref{fig:Lambdaresult}) and for all models with VLL triplets (the last two columns in Fig.~\ref{fig:Lambdaresult}). At small Yukawa couplings
$\lambdamax$, the singlet models become similar
to the SM, while the triplet models can be stable up to the Planck scale for
$\lambdamax \lesssim 0.25$
  thanks to the triplet gauge
interactions.
At larger $\lambdamax$ all models are less stable than the SM, and at
very large $\lambdamax$ some models show discontinuities due to effects exemplified in Fig.~\ref{fig:PhaseChange-Illustration}.  

For ease of use for further phenomenological analyses, we have also
approximated the resulting instability scales $\Lambda$ as a function of
$\lambdamax$ for each model class and each choice of nonzero Yukawas
and mass scale using simple parametric fits. The fit function is
given by 
\begin{align}
	\log_{10}(\Lambda/\SI{1}{GeV}) = a_0 + \frac{a_1-a_0}{1+\left(a_2 \lambdamax\right)^4}
        ,
\end{align}
such that the fit function interpolates between $a_1$ and $a_0$ when
$\lambdamax$ varies from $0$ to $\infty$. The fit parameters $a_0$,
$a_1$ and $a_2$ for the various cases are given in Table~\ref{tab:fitParameters}. 
\begin{table}[t]
	\centering
	\def\arraystretch{1.2}
	\begin{tabular}[t]{cc|cc|cc|cc}
		\multicolumn{2}{c|}{Parameters} & \multicolumn{2}{c|}{$a_0$} & \multicolumn{2}{c|}{$a_1$} & \multicolumn{2}{c}{$a_2$} \\ \hline
		$\lambda_\text{max}$ & $M$ & singlet & triplet & singlet & triplet\ \protect\footnotemark & singlet & triplet\\ \hline
		$\lambar$ & $\hphantom{0}2$ TeV  & 4.1 & 4.2 & 12 & 16 & 1.6 & 2.5\\
		$\lambar$ & $10$ TeV & 4.4 & 5.1 & 12 & 16 & 1.5 & 2.4\\
		$\lambda^{L,E}_2 \, \lambda\, \lambar$ & $\hphantom{0}2$ TeV  & 4.2 & 4.5 & 12 & 16 & 2.6 & 3.9\\
		$\lambda^{L,E}_2\, \lambda\, \lambar$ & $10$ TeV & 4.6 & 5.2 & 12 & 16 & 2.5 & 3.7\\
	\end{tabular}
	\caption{Fit parameters for the estimate of the instability scale $\Lambda(\lambdamax)$. The first column shows the VLL couplings set to $\lambda_\text{max}$ at $Q=M_Z$.
          The VLL masses are set to $M_L = M_E \equiv M$. The columns show the values for models with either a VLL singlet and triplet.}
	\label{tab:fitParameters}
\end{table}
	\footnotetext{For the triplet models, the value $a_1$  is chosen such that the fit shows good agreement for $\lambdamax \gtrsim 0.25$. For $\lambdamax \lesssim 0.25$ the triplet models typically do not develop a second minimum and the fit underestimates the instability scale.}
	
The choice of this function is motivated by the solution of the one-loop RGE for the Higgs quartic coupling $\lambda_\Phi$ and resulting
value $t(\lambdamax)$ of the implicit curve $\lambda_\Phi(\lambdamax, y_t, t(\lambdamax)) = 0$. 
For this, a simplified RGE system is solved in which of the SM Yukawas only the top Yukawa
coupling is kept nonzero and constant, of the VLL couplings only
$\lambar$ is kept nonzero and the RGE for $\lambda_\Phi$ is simplified
by neglecting terms proportional to the gauge couplings as well as
terms $\O{\left([\lambda_\Phi-\lambda_\Phi(\mu_0)]^2\right)}$. 

Within the uncertainty bands of the scan, to a good approximation the fit parameters
$a_1$ and $a_2$ are independent of the mass scale, only depending on
the singlet/triplet model class as well as the number of nonzero Yukawa
 couplings. In contrast, the parameter $a_0$ also depends on the mass
scale.

The fit shows rather good agreement in regions in which the
instability scale $\Lambda$ is differentiable in
$\lambdamax$ and no discontinuous transitions occur and where the instability
scale is orders of magnitude larger than the VLL mass scale. In
regions with larger couplings, where the instability scale is closer
to the VLL mass scale, 
the fit is less accurate but mostly underestimates the instability
scale. Hence, using the fit provides a conservative estimate of the maximum allowed
$\lambdamax$ up to $\lambdamax \approx 1.5$, beyond which
the computations can become unstable. 

%% file: Conclusions.tex
\section{Summary and Conclusions}\label{sec:Conclusion}

The six VLL models with seesaw-like contributions to lepton masses
lead to rich phenomenology in the lepton and Higgs sector. Past
investigations have already revealed the potential importance of one- and two-loop
corrections \cite{Mohling:2024qvk,Mohling:2026fqu}.

In the present paper we carried out systematic higher-order analyses
in all six VLL models. As a first step, renormalisation schemes
with on-shell mass conditions were defined. The setup resembles the
ones for the MSSM chargino/neutralino sectors, but it involves up to
$5\times5$ mixing matrices, and it turns out to be possible to define
all SM masses and (depending on models and scheme variants) most heavy masses as on-shell masses. 
The remaining parameters are taken as
the fundamental gauge-invariant couplings of the Lagrangian,
renormalised in the \MSbar-scheme. The advantages of the schemes include
a clean physical interpretation and the straightforward
practical implementation of computations where it is desirable to fix
lepton masses to their observed values.

We also determined the two-loop RGE $\beta$-functions appropriate for
the proposed schemes after converting the fundamental parameters to
running parameters in the \MSbar-scheme. These
results differ from generic results in the
literature because $\beta$-functions for unphysical parameters are
eliminated via 
field redefinitions, leading to additional terms for the remaining 
$\beta$-functions.

The setup was used to study the effective Higgs potential at next-to-leading log precision, 
including one-loop fixed-order and two-loop running effects.
With these improvements, our uncertainty estimates show that
robust conclusions can be drawn. Several qualitative features are
worth noting.

First, in VLL models with SU$(2)_L$ triplets, the electroweak vacuum tends
to be stabilised. As long as the BSM Yukawa couplings are smaller than
around $0.25$, the effective potential does not develop a second
minimum up to and beyond the Planck scale. In contrast, even for
vanishing BSM couplings, the VLL models
with SU$(2)_L$ singlets do not lead to a significant stabilisation.

If the BSM Yukawa couplings are larger, the VLL fermion loops
decrease the effective quartic Higgs coupling in all six models, and
the electroweak vacuum can become unstable. Depending on the values of
the couplings, the models show a rich behaviour, including discontinuous transitions, as illustrated
in Figs.~\ref{fig:lambdaEffplots}. The study shows that the
instability scale is mainly determined by the maximum of the BSM Yukawa couplings
$\lambdamax$, the number of Yukawa couplings close to this value and whether the model contains a singlet or triplet VLL.

The resulting instability scale $\Lambda$ was then studied in all
models as a function of $\lambdamax$ for various patterns of couplings
and masses. There is a complementarity to the parameter dependence of
EWPOs studied in the literature, implying that vacuum stability
constraints are particularly relevant in the case of VLL
masses above 2 TeV. In this case, coupling values allowed by
EWPOs decrease the instability scale from its SM value around
$10^{11}$~GeV down to $10^{5\ldots8}$~GeV in our examples. We further
derived simple numerical fit 
formulas for the behaviour of the instability scale. They provide
conservative estimates of $\Lambda$ in continuous regions with
$\lambdamax\lesssim1.5$ and can aid
future phenomenological analyses.

In summary, the setup of the on-shell renormalisation schemes and the
redundancy-free $\beta$-functions allows precise computations in VLL models,
keeping the physical lepton masses fixed at their observed values
despite the strongly modified mass generation mechanism. Similar to
earlier examples for Higgs decays and lepton dipole moments
\cite{Mohling:2024qvk,Mohling:2026fqu}, 
the quantum corrections are sizeable and important for robust predictions.
The same setup can also be used to compute further observables in the
future, such as flavour- and/or CP-violating Higgs or Z decays in
order to further constrain the parameter space of VLL models.

%% file: Appendix.tex
\appendix
\section{Renormalisation constants and coefficient matrices} \label{app:RC-mat}

\paragraph{$L\oplus E$}

\begin{align}
	\delta \mathcal{Y}^h_e = 
	\begin{pmatrix}
		\delta y_i \delta_{ij} & \delta\lambda^E_i & 0\\
		\delta \lambda^L_j & \delta \lambda & 0 \\
		0 & 0 & \delta \lambar
	\end{pmatrix}
	\qquad
	\delta \mathcal{M}^D_e = \begin{pmatrix}
		0 & 0 & 0 \\
		0 & 0 & \delta M_L \\
		0 & \delta M_E & 0
	\end{pmatrix}
	\qquad
	\delta \mathcal{M}^D_\nu = \begin{pmatrix}
		0 & 0 \\
		0 & \delta M_L
	\end{pmatrix}
\end{align}

\begin{align}
	\bm{\kappa}_\text{OS1} = \begin{pmatrix}
		U^{e*}_{L,11} U^{e}_{R,11} & U^{e*}_{L,21} U^{e}_{R,21} & U^{e*}_{L,31} U^{e}_{R,31} & U^{e*}_{L,41} U^{e}_{R,41} & U^{e*}_{L,51} U^{e}_{R,41} & U^{e*}_{L,41} U^{e}_{R,51} \\
		U^{e*}_{L,12} U^{e}_{R,12} & U^{e*}_{L,22} U^{e}_{R,22} & U^{e*}_{L,32} U^{e}_{R,32} & U^{e*}_{L,42} U^{e}_{R,42} & U^{e*}_{L,52} U^{e}_{R,42} & U^{e*}_{L,42} U^{e}_{R,52} \\
		U^{e*}_{L,13} U^{e}_{R,13} & U^{e*}_{L,23} U^{e}_{R,23} & U^{e*}_{L,33} U^{e}_{R,33} & U^{e*}_{L,43} U^{e}_{R,43} & U^{e*}_{L,53} U^{e}_{R,43} & U^{e*}_{L,43} U^{e}_{R,53} \\
		U^{e*}_{L,14} U^{e}_{R,14} & U^{e*}_{L,24} U^{e}_{R,24} & U^{e*}_{L,34} U^{e}_{R,34} & U^{e*}_{L,44} U^{e}_{R,44} & U^{e*}_{L,54} U^{e}_{R,44} & U^{e*}_{L,44} U^{e}_{R,54} \\
		U^{e*}_{L,15} U^{e}_{R,15} & U^{e*}_{L,25} U^{e}_{R,25} & U^{e*}_{L,35} U^{e}_{R,35} & U^{e*}_{L,45} U^{e}_{R,45} & U^{e*}_{L,55} U^{e}_{R,45} & U^{e*}_{L,45} U^{e}_{R,55} \\
		0 & 0 & 0 & 0 & 0 & 1
	\end{pmatrix}
\end{align}

\paragraph{$L\oplus N$ }

\begin{subequations}
	\begin{alignat}{3}
		\delta \mathcal{Y}^h_e &= \begin{pmatrix}
			\delta y_i \delta_{ij} & 0\\
			\delta \lambda^L_j & 0
		\end{pmatrix} &&
		\qquad
		\delta \mathcal{M}^D_e &&= \begin{pmatrix}
			0 & 0 \\
			0 & \delta M_L
		\end{pmatrix} \\
		\delta \mathcal{Y}^h_\nu &= \begin{pmatrix}
			0 & 0 & \delta \lambda^N_i\\
			0 & 0 & \delta \lambda \\
			0 & \delta\lambar & 0
		\end{pmatrix} &&
		\qquad
		\delta \mathcal{M}^D_\nu &&= \begin{pmatrix}
			0 & 0 & 0 \\
			0 & \delta M_L & 0\\
			0 & 0 & \delta M_N
		\end{pmatrix}
	\end{alignat}
\end{subequations}
\begin{align}
	\bm{\kappa}_\text{OS1} = \begin{pmatrix}
		U^{e*}_{L,11} U^{e}_{R,11} & U^{e*}_{L,21} U^{e}_{R,21} & U^{e*}_{L,31} U^{e}_{R,31} & 0 & 0 & U^{e*}_{L,41} U^{e}_{R,41} \\
		U^{e*}_{L,12} U^{e}_{R,12} & U^{e*}_{L,22} U^{e}_{R,22} & U^{e*}_{L,32} U^{e}_{R,32} & 0 & 0 & U^{e*}_{L,42} U^{e}_{R,42} \\
		U^{e*}_{L,13} U^{e}_{R,13} & U^{e*}_{L,23} U^{e}_{R,23} & U^{e*}_{L,33} U^{e}_{R,33} & 0 & 0 & U^{e*}_{L,43} U^{e}_{R,43} \\
		U^{e*}_{L,14} U^{e}_{R,14} & U^{e*}_{L,24} U^{e}_{R,24} & U^{e*}_{L,34} U^{e}_{R,34} & 0 & 0 & U^{e*}_{L,44} U^{e}_{R,44} \\
		0 & 0 & 0 & U^{\nu*}_{L,44} U^{\nu}_{R,54} & U^{\nu*}_{L,54} U^{\nu}_{R,54} & U^{\nu*}_{L,44} U^{\nu}_{R,44} \\
		0 & 0 & 0 & U^{\nu*}_{L,45} U^{\nu}_{R,55} & U^{\nu*}_{L,55} U^{\nu}_{R,55} & U^{\nu*}_{L,45} U^{\nu}_{R,45}
	\end{pmatrix}
\end{align}

\paragraph{$L_{\frac32}\oplus E$} The matrices are the same as for $L\oplus E$ except for $\delta\mathcal{M}^D_\nu = 0$ and $\delta\mathcal{M}^D_\rho = \delta M_L$.

\paragraph{$L\oplus E^a$} $\delta\mathcal{M}^D_\rho = \delta M_E$ and 
\begin{subequations}
	\begin{alignat}{3}
		\delta \mathcal{Y}^h_e &= 
		\begin{pmatrix}
			\delta y_i \delta_{ij} & -\delta\lambda^E_i & 0\\
			\delta \lambda^L_j & -\delta \lambda & 0 \\
			0 & 0 & -\delta \lambar
		\end{pmatrix} &&
		\qquad
		\delta \mathcal{M}^D_e &&= \begin{pmatrix}
			0 & 0 & 0 \\
			0 & 0 & \delta M_L \\
			0 & \delta M_E & 0
		\end{pmatrix} \\
		\delta \mathcal{Y}^h_\nu &= \sqrt{2}
		\begin{pmatrix}
			0 & \delta\lambda^E_i\\
			0 & \delta \lambda \\
			\delta \lambar & 0
		\end{pmatrix} &&
		\qquad
		\delta \mathcal{M}^D_\nu &&= \begin{pmatrix}
			0 & 0 \\
			\delta M_L & 0 \\
			0 & \delta M_E
		\end{pmatrix}
	\end{alignat}
\end{subequations}

\begin{align}
	\bm{\kappa}_\text{OS1} = 
	\begin{pmatrix}
		U^{e*}_{L,11} U^{e}_{R,11} & U^{e*}_{L,21} U^{e}_{R,21} & U^{e*}_{L,31} U^{e}_{R,31} & -(U^{e*}_{L,41} U^{e}_{R,41}) & -(U^{e*}_{L,51} U^{e}_{R,51}) & U^{e*}_{L,51} U^{e}_{R,41} & U^{e*}_{L,41} U^{e}_{R,51} \\
		U^{e*}_{L,12} U^{e}_{R,12} & U^{e*}_{L,22} U^{e}_{R,22} & U^{e*}_{L,32} U^{e}_{R,32} & -(U^{e*}_{L,42} U^{e}_{R,42}) & -(U^{e*}_{L,52} U^{e}_{R,52}) & U^{e*}_{L,52} U^{e}_{R,42} & U^{e*}_{L,42} U^{e}_{R,52} \\
		U^{e*}_{L,13} U^{e}_{R,13} & U^{e*}_{L,23} U^{e}_{R,23} & U^{e*}_{L,33} U^{e}_{R,33} & -(U^{e*}_{L,43} U^{e}_{R,43}) & -(U^{e*}_{L,53} U^{e}_{R,53}) & U^{e*}_{L,53} U^{e}_{R,43} & U^{e*}_{L,43} U^{e}_{R,53} \\
		U^{e*}_{L,14} U^{e}_{R,14} & U^{e*}_{L,24} U^{e}_{R,24} & U^{e*}_{L,34} U^{e}_{R,34} & -(U^{e*}_{L,44} U^{e}_{R,44}) & -(U^{e*}_{L,54} U^{e}_{R,54}) & U^{e*}_{L,54} U^{e}_{R,44} & U^{e*}_{L,44} U^{e}_{R,54} \\
		U^{e*}_{L,15} U^{e}_{R,15} & U^{e*}_{L,25} U^{e}_{R,25} & U^{e*}_{L,35} U^{e}_{R,35} & -(U^{e*}_{L,45} U^{e}_{R,45}) & -(U^{e*}_{L,55} U^{e}_{R,55}) & U^{e*}_{L,55} U^{e}_{R,45} & U^{e*}_{L,45} U^{e}_{R,55} \\
		0 & 0 & 0 & \sqrt2 U^{\nu*}_{L,44} U^{\nu}_{R,54} & \sqrt2 U^{\nu*}_{L,54} U^{\nu}_{R,44} & U^{\nu*}_{L,54} U^{\nu}_{R,54} & U^{\nu*}_{L,44} U^{\nu}_{R,44} \\
		0 & 0 & 0 & \sqrt2 U^{\nu*}_{L,45} U^{\nu}_{R,55} & \sqrt2 U^{\nu*}_{L,55} U^{\nu}_{R,45} & U^{\nu*}_{L,55} U^{\nu}_{R,55} & U^{\nu*}_{L,45} U^{\nu}_{R,45}
	\end{pmatrix}
\end{align}

\paragraph{$L\oplus N^a$} $\delta\mathcal{M}^D_\chi = \delta M_N$ and 
\begin{subequations}
	\begin{alignat}{3}
		\delta \mathcal{Y}^h_e &= 
		\begin{pmatrix}
			\delta y_i \delta_{ij} & \sqrt{2}\delta\lambda^E_i & 0\\
			\delta \lambda^L_j & \sqrt{2}\,\delta \lambda & 0 \\
			0 & 0 & \sqrt{2}\delta \lambar
		\end{pmatrix} &&
		\qquad
		\delta \mathcal{M}^D_e &&= \begin{pmatrix}
			0 & 0 & 0 \\
			0 & 0 & \delta M_L \\
			0 & \delta M_N & 0
		\end{pmatrix} \\
		\delta \mathcal{Y}^h_\nu &= 
		\begin{pmatrix}
			0 & \delta\lambda^E_i\\
			0 & \delta \lambda \\
			\delta \lambar & 0
		\end{pmatrix} &&
		\qquad
		\delta \mathcal{M}^D_\nu &&= \begin{pmatrix}
			0 & 0 \\
			\delta M_L & 0 \\
			0 & \delta M_N
		\end{pmatrix}
	\end{alignat}
\end{subequations}

\begin{align}
	\bm{\kappa}_\text{OS1} = 
	\begin{pmatrix}
		U^{e*}_{L,11} U^{e}_{R,11} & U^{e*}_{L,21} U^{e}_{R,21} & U^{e*}_{L,31} U^{e}_{R,31} & \sqrt2 U^{e*}_{L,41} U^{e}_{R,41} & \sqrt2 U^{e*}_{L,51} U^{e}_{R,51} & U^{e*}_{L,51} U^{e}_{R,41} & U^{e*}_{L,41} U^{e}_{R,51} \\
		U^{e*}_{L,12} U^{e}_{R,12} & U^{e*}_{L,22} U^{e}_{R,22} & U^{e*}_{L,32} U^{e}_{R,32} & \sqrt2 U^{e*}_{L,42} U^{e}_{R,42} & \sqrt2 U^{e*}_{L,52} U^{e}_{R,52} & U^{e*}_{L,52} U^{e}_{R,42} & U^{e*}_{L,42} U^{e}_{R,52} \\
		U^{e*}_{L,13} U^{e}_{R,13} & U^{e*}_{L,23} U^{e}_{R,23} & U^{e*}_{L,33} U^{e}_{R,33} & \sqrt2 U^{e*}_{L,43} U^{e}_{R,43} & \sqrt2 U^{e*}_{L,53} U^{e}_{R,53} & U^{e*}_{L,53} U^{e}_{R,43} & U^{e*}_{L,43} U^{e}_{R,53} \\
		U^{e*}_{L,14} U^{e}_{R,14} & U^{e*}_{L,24} U^{e}_{R,24} & U^{e*}_{L,34} U^{e}_{R,34} & \sqrt2 U^{e*}_{L,44} U^{e}_{R,44} & \sqrt2 U^{e*}_{L,54} U^{e}_{R,54} & U^{e*}_{L,54} U^{e}_{R,44} & U^{e*}_{L,44} U^{e}_{R,54} \\
		U^{e*}_{L,15} U^{e}_{R,15} & U^{e*}_{L,25} U^{e}_{R,25} & U^{e*}_{L,35} U^{e}_{R,35} & \sqrt2 U^{e*}_{L,45} U^{e}_{R,45} & \sqrt2 U^{e*}_{L,55} U^{e}_{R,55} & U^{e*}_{L,55} U^{e}_{R,45} & U^{e*}_{L,45} U^{e}_{R,55} \\
		0 & 0 & 0 & U^{\nu*}_{L,44} U^{\nu}_{R,54} & U^{\nu*}_{L,54} U^{\nu}_{R,44} & U^{\nu*}_{L,54} U^{\nu}_{R,54} & U^{\nu*}_{L,44} U^{\nu}_{R,44} \\
		0 & 0 & 0 & U^{\nu*}_{L,45} U^{\nu}_{R,55} & U^{\nu*}_{L,55} U^{\nu}_{R,45} & U^{\nu*}_{L,55} U^{\nu}_{R,55} & U^{\nu*}_{L,45} U^{\nu}_{R,45}
	\end{pmatrix}
\end{align}

\paragraph{$L_\frac32\oplus E^a$}

\begin{subequations}
	\begin{alignat}{3}
		\delta \mathcal{Y}^h_e &= 
		\begin{pmatrix}
			\delta y_i \delta_{ij} & -\delta\lambda^E_i & 0\\
			\delta \lambda^L_j & \delta \lambda & 0 \\
			0 & 0 & \delta \lambar
		\end{pmatrix} &&
		\qquad
		\delta \mathcal{M}^D_e &&= \begin{pmatrix}
			0 & 0 & 0 \\
			0 & 0 & \delta M_L \\
			0 & \delta M_E & 0
		\end{pmatrix} \\
		\delta \mathcal{Y}^h_\nu &= 
		\begin{pmatrix}
			\sqrt{2} \delta\lambda^E_i\\
			0
		\end{pmatrix} &&
		\qquad
		\delta \mathcal{M}^D_\nu &&= \begin{pmatrix}
			0 \\
			\delta M_E
		\end{pmatrix} \\
		\delta \mathcal{Y}^h_\rho &=  \sqrt{2}
		\begin{pmatrix}
			0 & \delta \lambda \\
			\delta \lambar & 0
		\end{pmatrix} &&
		\qquad
		\delta \mathcal{M}^D_\rho &&= \begin{pmatrix}
			\delta M_L & 0 \\
			0 & \delta M_E
		\end{pmatrix}
	\end{alignat}
\end{subequations}

\begin{align}
	\bm{\kappa}_\text{OS1} = 
	\begin{pmatrix}
		U^{e*}_{L,11} U^{e}_{R,11} & U^{e*}_{L,21} U^{e}_{R,21} & U^{e*}_{L,31} U^{e}_{R,31} & U^{e*}_{L,41} U^{e}_{R,41} & U^{e*}_{L,51} U^{e}_{R,51} & U^{e*}_{L,51} U^{e}_{R,41} & U^{e*}_{L,41} U^{e}_{R,51} \\
		U^{e*}_{L,12} U^{e}_{R,12} & U^{e*}_{L,22} U^{e}_{R,22} & U^{e*}_{L,32} U^{e}_{R,32} & U^{e*}_{L,42} U^{e}_{R,42} & U^{e*}_{L,52} U^{e}_{R,52} & U^{e*}_{L,52} U^{e}_{R,42} & U^{e*}_{L,42} U^{e}_{R,52} \\
		U^{e*}_{L,13} U^{e}_{R,13} & U^{e*}_{L,23} U^{e}_{R,23} & U^{e*}_{L,33} U^{e}_{R,33} & U^{e*}_{L,43} U^{e}_{R,43} & U^{e*}_{L,53} U^{e}_{R,53} & U^{e*}_{L,53} U^{e}_{R,43} & U^{e*}_{L,43} U^{e}_{R,53} \\
		U^{e*}_{L,14} U^{e}_{R,14} & U^{e*}_{L,24} U^{e}_{R,24} & U^{e*}_{L,34} U^{e}_{R,34} & U^{e*}_{L,44} U^{e}_{R,44} & U^{e*}_{L,54} U^{e}_{R,54} & U^{e*}_{L,54} U^{e}_{R,44} & U^{e*}_{L,44} U^{e}_{R,54} \\
		U^{e*}_{L,15} U^{e}_{R,15} & U^{e*}_{L,25} U^{e}_{R,25} & U^{e*}_{L,35} U^{e}_{R,35} & U^{e*}_{L,45} U^{e}_{R,45} & U^{e*}_{L,55} U^{e}_{R,55} & U^{e*}_{L,55} U^{e}_{R,45} & U^{e*}_{L,45} U^{e}_{R,55} \\
		0 & 0 & 0 & \sqrt2 U^{\rho*}_{L,11} U^{\rho}_{R,21} & \sqrt2 U^{\rho*}_{L,21} U^{\rho}_{R,11} & U^{\rho*}_{L,21} U^{\rho}_{R,21} & U^{\rho*}_{L,11} U^{\rho}_{R,11} \\
		0 & 0 & 0 & \sqrt2 U^{\rho*}_{L,12} U^{\rho}_{R,22} & \sqrt2 U^{\rho*}_{L,22} U^{\rho}_{R,12} & U^{\rho*}_{L,22} U^{\rho}_{R,22} & U^{\rho*}_{L,12} U^{\rho}_{R,12}
	\end{pmatrix}
\end{align}